\documentclass[%
a4paper,							
11pt,								
bibliography=totoc,						
abstracton,					
]
{scrartcl}

\usepackage[a4paper,left=3.0cm,right=2.5cm, top=2.5cm, bottom=3.0cm]{geometry}
\usepackage[headsepline=.4pt,footsepline=.4pt,automark,autooneside=false,]{scrlayer-scrpage}
\clearpairofpagestyles
\automark[subsection]{section} 
\automark*[subsubsection]{subsection}
\ihead{\scriptsize\rightmark} 
\usepackage[ngerman, english]{babel}
\usepackage[T1]{fontenc}
\usepackage[utf8]{inputenc}
\usepackage{lmodern} 
\usepackage[onehalfspacing]{setspace} 

\usepackage{xspace}
\usepackage{calc}

\usepackage{amsmath}
\usepackage{amssymb}
\usepackage{amsthm,thmtools}

\usepackage{mathtools}
\usepackage{bbm}
\usepackage{bm}
\usepackage[mathscr]{eucal}
\begingroup
    \makeatletter
    \@for\theoremstyle:=definition,remark,plain\do{%
        \expandafter\g@addto@macro\csname th@\theoremstyle\endcsname{%
            \addtolength\thm@preskip\parskip
            }%
        }
\endgroup

\usepackage{chngcntr}
\counterwithin*{equation}{section}

\theoremstyle{plain}
\newtheorem{theorem}{Theorem}[section]

\theoremstyle{definition}

\theoremstyle{remark}
\newtheorem{remark}[theorem]{Remark}
\theoremstyle{plain}

\theoremstyle{plain}

\theoremstyle{remark}

\usepackage{tabularx}
\usepackage{booktabs}
\usepackage{multirow}
\usepackage{multicol}
\usepackage{pdflscape}

\usepackage{diagbox}

\usepackage[flushleft]{paralist}

\usepackage{color}

\usepackage{graphicx}
\usepackage{tikz}
\usetikzlibrary{calc,spy,backgrounds,patterns,arrows,intersections}
\usepackage[final]{showkeys}

\usepackage{fancyvrb}

\usepackage{float}
\usepackage[section]{placeins}
\usepackage[format=plain,tableposition=top]{caption}
\usepackage[position=top]{subcaption}
\usepackage{algorithm2e}
\usepackage{csquotes} 

\usepackage{titleref}
\usepackage{hyperref}
\usepackage{cleveref}
\hypersetup{
    colorlinks,
	linkcolor={red},
    citecolor={blue},
    urlcolor={blue}
}

\renewcommand{\tilde}[1]{\widetilde{#1}}
\renewcommand{\P}{\mathbb{P}}

\newcommand{\R}{\mathbb{R}}
\newcommand{\Q}{\mathbb{Q}}
\newcommand{\N}{\mathbb{N}}
\newcommand{\1}{\mathbbm{1}}

\newcommand\Lc{\mathscr{L}}
\newcommand\Qc{\mathscr{Q}}

\newcommand\dD{\Lc}
\newcommand\dDD{\Qc}

\newcommand{\lieAlgebra}{\mathfrak{g}}
\newcommand{\lieBasis}{\mathcal{E}}

\newcommand{\ad}{\mathrm{ad}}

\newcommand{\norm}[1]{\left\|#1\right\|}
\newcommand{\abs}[1]{\left|#1\right|}
\newcommand{\vectorize}{\mathrm{vec}}

\newcommand{\rFormat}[1]{\text{\textbf{#1}}\xspace}
\newcommand{\rProcess}{R}

\newcommand{\rAdjusted}{\rProcess^{\mathrm{A}}}

\newcommand{\rSDE}{\rProcess^{\mathrm{SDE}}}

\newcommand{\rGEM}{\rProcess^{\mathrm{gEM}}}
\newcommand{\rCohort}{\rProcess^{\mathrm{cohort}}}
\newcommand{\rRec}{\rProcess^{\mathrm{Rec}}}

\newcommand{\PD}{\mathrm{PD}}

\newcommand{\generator}{A}

\newcommand{\girsanov}{L}
\newcommand{\ratingProcess}{X}

\newcommand{\exposure}{\mathrm{V}}

\newcommand{\lgd}{\mathrm{LGD}}

\newcommand{\CBVA}{\mathrm{BVA}}
\newcommand{\CDVA}{\mathrm{DVA}}
\newcommand{\CCVA}{\mathrm{CVA}}
\newcommand{\textCBVA}{$\CBVA$\xspace}
\newcommand{\textCDVA}{$\CDVA$\xspace}
\newcommand{\textCCVA}{$\CCVA$\xspace}
\newcommand{\textCXVA}{$\mathrm{XVA}$\xspace}
\newcommand{\collateral}{\mathrm{C}}

\newcolumntype{C}{X<{\centering}}

\newcommand{\CPU}{Intel(R) Core(TM) i7-8750H CPU @ 2.20\,GHz\xspace}
\newcommand{\RAM}{2x32\,GB (Dual Channel) Samsung SODIMM DDR4 RAM @ 2667 MHz\xspace}
\newcommand{\GPU}{NVIDIA GeForce RTX 2070 with Max-Q Design (8\,GB GDDR6 RAM)\xspace}
\newcommand{\OS}{Windows 10 Pro\xspace}

\SaveVerb{python}=(Intel-)Python 3.9=
\newcommand{\python}{\protect\UseVerb{python}\xspace}
\SaveVerb{tensorflow}=Tensorflow 2.8.0=
\newcommand{\tensorflow}{\protect\UseVerb{tensorflow}\xspace}
\SaveVerb{matlab}=Matlab 2022a=
\newcommand{\matlab}{\protect\UseVerb{matlab}\xspace}
\SaveVerb{ga}=ga=

\SaveVerb{fmincon}=fmincon=

\SaveVerb{fitdist}=fitdist=

\SaveVerb{lsqnonlin}=lsqnonlin=
\newcommand{\lsqnonlin}{\protect\UseVerb{lsqnonlin}\xspace}
\newcommand{\matlabGOtoolbox}{(Global) Optimization Toolbox\xspace}

\newcommand{\suffix}{-eps-converted-to.pdf}

\title{Rating Triggers for Collateral-Inclusive XVA via Machine Learning and SDEs on Lie Groups}
\author{Kevin Kamm\thanks{Dipartimento di Matematica, Universit\`a di Bologna, Bologna, Italy.
\textbf{e-mail}: kevin.kamm@unibo.it}
\and Michelle Muniz\thanks{Institute of Mathematical Modelling, Analysis and Computational Mathematics (IMACM), 
Chair of Applied and Computational Mathematics, Bergische Universität Wuppertal, Wuppertal, Germany.
\textbf{e-mail}: muniz@uni-wuppertal.de}
}

\begin{document}
\thispagestyle{empty}\pagenumbering{roman}
\maketitle
\renewcommand{\thefootnote}{\Roman{footnote}}
\renewcommand{\thefootnote}{\arabic{footnote}}
\begin{abstract}
In this paper, we model the rating process of an entity by using a geometrical approach. We model rating transitions as an SDE on a Lie group. Specifically, we focus on calibrating the model to both historical data (rating transition matrices) and market data (CDS quotes) and compare the most popular choices of changes of measure to switch from the historical probability to the risk-neutral one. For this, we show how the classical Girsanov theorem can be applied in the Lie group setting. Moreover, we overcome some of the imperfections of rating matrices published by rating agencies, which are computed with the cohort method, by using a novel Deep Learning approach. This leads to an improvement of the entire scheme and makes the model more robust for applications. We apply our model to compute bilateral credit and debit valuation adjustments of a netting set under a CSA with thresholds depending on ratings of the two parties.
\end{abstract}
\noindent
\textbf{Acknowledgements:}
This project has received funding from the European Union’s Horizon 2020 research and innovation programme
under the Marie Sklodowska-Curie grant agreement No 813261 and is part of the ABC-EU-XVA
project. We would like to thank Balint Negyesi for many inspiring discussions about machine learning techniques.\\\noindent
\textbf{Code availability:}
The code and data sets to produce the numerical experiments are available at
\url{https://github.com/kevinkamm/StochasticCohort}.
\newpage
\pagestyle{scrheadings}\ihead{\scriptsize\rightmark}\pagenumbering{arabic}
\section{Introduction}\label{sec:introduction}
In this paper, we extend the rating model found in \cite{KM2022} to allow for data under the historical and risk-neutral measure by applying Girsanov's theorem in a Lie group setting. Moreover, we discuss rating matrices computed by the so-called \emph{cohort method} and show how the Autoencoder of a TimeGAN can be utilized to recover a valid rating matrix. After calibrating the rating transition model to the historical and risk-neutral data, we apply it to compute credit and debit valuation adjustments (hereafter referred to as \textCCVA and \textCDVA)  
of a portfolio of trades between two parties having signed a collateral
agreement dependent on ratings like in \cite{K2022}.

Let us briefly recall for the convenience of the reader the concept of ratings.
A rating is an indicator of the creditworthiness of an entity. A high rating associates less risk to an entity to not fulfill its financial obligations and a low rating a high risk.
Ratings are usually denoted by letters
$\rFormat{A}$, $\rFormat{B}$, \dots, $\rFormat{D}$, where $\rFormat{A}$ denotes the best rating and
$\rFormat{D}$ denotes the worst rating. The rating $\rFormat{D}$ is special. It means that an entity has defaulted, i.e.\ it cannot fulfill its financial obligation towards a contracting party. In this paper, we use the terms default and bankruptcy of an entity synonymously, implying that a defaulted company cannot recover from this state.

To keep this presentation as simple as possible, we consider only four different ratings: \rFormat{A}, \rFormat{B}, \rFormat{C}, \rFormat{D} ordered from best to worst rating and identify them by integers $\left\{1,2,\dots,K\right\}$, whenever it is more convenient. But it is straightforward to use more ratings.

For our main application to collateralized \textCXVA, it is important to model the rating changes of an individual entity or an entire sector on a continuous time scale. 
This can be done in two different ways. On the one hand, one can define a process $X_t$, which tells us at each time and trajectory the current rating of a company. The natural state-space of these processes is therefore discrete and the time axis is continuous.
On the other hand, one can model the transition probabilities $R_t$ of a sector at each point in time and derive a rating process using these transition probabilities. The state-space of this type of model is then a matrix whose entries are the probability of transitioning from one rating to another starting at an initial time $t_0$ (usually today) till a future time $t$. An example of such a $t-t_0$ rating matrix is given in \Cref{tab:ratingMatrix1}.
\begin{table}[h]
    \centering
    \begin{tabular}{|c|*{4}{c}|} 
        \hline
        \diagbox[]{From}{To}  & \rFormat{A} & \rFormat{B} & \rFormat{C} & \rFormat{D}\\
        \hline
        \rFormat{A} & 0.9395 & 0.0566 & 0.0037 & 2.7804e-04\\ 
        \rFormat{B} & 0.0092 & 0.9680 & 0.0211 & 0.0017\\ 
        \rFormat{C} & 6.2064e-04 & 0.0440 & 0.8154 & 0.1400\\ 
        \rFormat{D} & 0 & 0 & 0 & 1\\
        \hline
    \end{tabular}
    \caption{Example of a one year rating transition matrix.}
    \label{tab:ratingMatrix1}
\end{table}
We can see that the individual rows sum up to one, meaning that all rows are valid probability distributions. These type of matrices are called stochastic for this reason. The last row corresponds to our idealized assumption that a defaulted entity cannot recover, i.e.\ the default state is absorbing.
Rating agencies publish these type of matrices usually once a year for a few time frames. Short-term rating matrices are usually published with time frames of $1,3,6,12$ months and long-term rating matrices
with time frames of $1,2,3,5,10$ years.
We see a lot of uncertainty in the historical data published by the agencies increasing with larger time frames, which we will discuss in more detail in \Cref{sec:LieCalibrationP}.

For a review of the relevant literature we refer the reader to \cite{K2022} and \cite{KM2022}.

In this paper, we extend the results from \cite{KM2022} by showing how to apply Girsanov's theorem to take market data under the risk-neutral measure, as well as historical data in terms of rating matrices into account.

We will use cohort rating matrices under the historical measure and propose a novel method using a Deep Neural Network (DNN) to deal with their imperfections due to withdrawals.
Moreover, we will take the uncertainty of this reconstruction into account to model a stochastic rating transition model in order to mitigate the effects from the choice of our proposed reconstruction method. To the best of our knowledge, this is the first paper which is proposing a stochastic reconstruction method for cohort rating matrices and is of value for pricing \textCXVA with rating triggers in a robust way.

The paper is structured as follows: In \Cref{sec:ratingHistoricalData}, we will explain the data we use for calibrating our rating transition model. The section is divided into two parts. In \Cref{sec:ratingCohort}, we explain how to remove the imperfections from a cohort rating matrix due to withdrawals by using a Deep Learning approach. This is followed by a brief reminder of Credit Default Swap quotes in \Cref{sec:defaultProb}.
After that, in \Cref{sec:SDELieGroup} we give a gentle introduction to matrix Lie groups and notice that the stochastic matrices are a subgroup of a matrix Lie group. We show how to utilize this framework to model rating transition matrices by a stochastic process and apply Girsanov's theorem to find the model dynamics under a new probability measure.
Afterwards, we do some numerical experiments in \Cref{sec:numerics} and define desirable properties of short-term rating matrices in \Cref{sec:errors}. The first step is to calibrate the rating process to the reconstruction of the cohort rating matrices while simultaneously accounting for the uncertainty coming from this reconstruction in \Cref{sec:LieCalibrationP}. In a second step, we calibrate the parameters of the change of measure to match the market default probabilities in our model in \Cref{sec:LieCalibrationQ}. In both cases, we assess the quality of the calibration by studying their distribution and the aforementioned rating properties.
After the calibration, we simulate rating processes with a nested stochastic simulation algorithm (SSA) in \Cref{sec:LieSimulation} and compute bilateral \textCCVA and \textCDVA with rating triggers in \Cref{sec:CXVA}.

\section{Historical and Market Data}\label{sec:ratingHistoricalData}
Rating agencies, such as S\&P, Moody's and Fitch are required by \enquote{Rule 17g-7 of the Securities Exchange Act of 1934}\footnote{Please visit \url{https://www.sec.gov/structureddata/rocr-publication-guide.html} for more details. Last accessed: 19.05.2022 12:23 CET.} to publish the history of rating changes for some entities. 
The data set can be downloaded from the websites of the rating agencies and consists of rating histories of individual entities in different sectors, e.g.\ financial and corporate. We will use the data set from S\&P with focus on the corporate sector. The data is structured as follows: for each entity it consists of a list of time stamps when a rating was changed or confirmed. Therefore, we can extract the historical ratings for each individual company for each day. 

There are two major methods how to process this data, the Aalen-Johansen estimation and the cohort method. For the Aalen-Johansen estimator, we refer the reader to \cite{KM2022} and \cite{Lando2002}.
\subsection{A Stochastic Reconstruction for the Cohort Method}\label{sec:ratingCohort}
The \emph{cohort method} is computed from the aforementioned rating histories as follows (cf. \cite[pp.~2\,ff. Equation 1]{Lando2002}):
Suppose, we have $N_i\in \N$ entities with a rating $i$ at the beginning of the year, $s=0$.
Now, we look at time $t>s$, e.g.\ the end of the same year, how these entities have changed their rating. We denote by $N_{ij}$ the number of entities, who transition from rating $i$ to rating $j$ and compute the corresponding transition probability as
\begin{align*}
	p_{ij}(s,t)\coloneqq \frac{N_{ij}}{N_i}.
\end{align*}
However, in practice it can happen for various reasons that $\sum_{j}^{}{N_{ij}}< N_i$, which implies that the transition matrices $P(s,t)\coloneqq \left(p_{ij}(s,t)\right)_{ij}$
do not have row sums equal to one. One reason why this is possible is the fact that companies can decide themselves that they do not want to be rated anymore by the rating agencies. We call this situation \emph{withdrawal} of an entity. 

Therefore, we see in the published rating matrices by the rating agencies that rows usually will not sum up to one and for large times only sum up to around $0.5$. 
A natural question that arises is how to repair these matrices, because for modelling rating transitions it is important that the rating matrices do not loose probability mass over time to use familiar concepts as Markov chains or take advantage of Lie group methods.

This problem is discussed in more details in \cite{Israel2001} and a simple reconstruction is recommended for small withdrawal rates. For larger withdrawal rates, we could not find a proper method in the literature and propose the method described in the following paragraph.

\paragraph*{Reconstruction by using a TimeGAN.} 
Let us explain how we use a trained TimeGAN (cf. \cite{Yoon2019}) to reconstruct a rating matrix computed by the cohort method with larger withdrawal rates. 
We refer to \cite{KM2022} for the training of the TimeGAN and the training dataset.

Let us denote by $\rCohort_{t}$ a rating matrix computed by the cohort method with possible row sums less than one
for $t$ equal to $1,3,6,12$ months. For simplicity, we will explain the idea of the reconstruction for a single rating matrix. Therefore, let us define $w_t\coloneqq 1-\left(\sum_{j=1}^{K}{\left(\rCohort_t\right)_{ij}}\right)_{i=1,\dots,K}$ as the vector of withdrawal rates for $\rCohort_t$.
Moreover, denote by $f\colon\R  \rightarrow \R^{K,K}_{>0}$ a function $f(W_t)$ depending a Brownian motion $W_t$ with values in the space of real-valued matrices with strictly positive values and $\bar{f}(W_t)\coloneqq \left(\sum_{j=1}^{K}{\left(f(W_t)\right)_{ij}}\right)_{i=1,\dots,K}$ the row sums of $f(W_t)$.

Now, define a new weight matrix $\nu_t^f\in [0,1]^{K,K}$ by
\begin{align*}
    \left(\nu_t^f\right)_{i,j} \coloneqq \frac{\left(f(W_t)\right)_{i,j} }{\left(\bar{f}(W_t)\right)_{i}}
\end{align*}
and set
\begin{align*}
    \left(\omega_t^f\right)_{i,j} \coloneqq \left(w_t\right)_{i} \cdot \left(\nu_t^f\right)_{i,j} 
\end{align*}
for $i,j=1,\dots,K$.

Then, the matrix
\begin{align*}
    {\rProcess}_t^{\mathrm{Rec},f} = \rCohort_t + \omega_t^f
\end{align*}
is guaranteed to have row sums equal to one for any function $f$.

We would like to learn the function $f$ given the cohort matrix $\rCohort_t$ by using a Deep Neural Network (DNN) such that the reconstructed rating matrix ${\rProcess}_t^{\mathrm{Rec},f}$ makes sense.

Our idea is to use the trained Autoencoder of the aforementioned TimeGAN to judge during the training phase of the new DNN whether the reconstructed rating matrix $\rRec_t$ is a proper rating matrix or not, without imposing further soft or hard constraints. 
Thus, let us denote by $e:\R^{K,K} \rightarrow \R^k$, $k\in\N$, the trained Embedder network of the TimeGAN's Autoencoder and 
by $r:\R^{k} \rightarrow \R^{K,K}$ the trained Recovery network.

We set the loss function for the training of our new DNN for $f$ as
\begin{align*}
    l\left({\rProcess}_t^{\mathrm{Rec},f}\right)\coloneqq \norm{{\rProcess}_t^{\mathrm{Rec},f} - (r\circ e)({\rProcess}_t^{\mathrm{Rec},f})}_{F},
\end{align*}
where $\norm{\cdot}_F$ denotes the Frobenius norm.

Since the TimeGAN was trained for a time-series of rating matrices, we train the DNN for $f$ by augmenting the previous idea such that it can take time-series data into account with the obvious changes. 

The heuristic idea of this approach is that the Autoencoder of the TimeGAN is able to recover the Aalen-Johansen counterpart of the cohort method, since the Embedder can match the principle values of the reconstructed rating matrix with the ones it learned during its training phase with Aalen-Johansen time-series data.

For the DNN for $f$ we use three deeply connected layers with $K^2$, $K^2\cdot N^2$, $K^2 \cdot N$ neurons. We also tried different numbers of neurons in the individual layers. Using more neurons lead to even better results but we found this configuration to be sufficient for our purposes. As an optimizer we used Adam (cf. \cite{Kingma2014}) with standard parameters in \tensorflow and $10000$ epochs. We leave further hyperparameter optimization of this approach to the reader, since the code will be publicly available.

To test the neural network, we used a time-series of rating matrices obtained by the Aalen-Johansen method from the historical rating dataset and subtracted at some entries a certain amount of probability mass to obtain a rating matrix with row sums less than one. This allows us to have a ground-truth matrix, which we can compare to the reconstruction obtained from the DNN.

In \Cref{tab:Rec}, we only show the one-year rating matrices, since the results for the $1,3,6$ month rating matrices are similar. 
In \Cref{tab:Rec_groundtruth}, we can see an example of an one-year rating matrix computed with the Aalen-Johansen method, in \Cref{tab:Rec_cohort} a rating matrix with row sums less than one, which we will refer to as the cohort matrix from now on, and the reconstructed rating matrix obtained after the training of the DNN in \Cref{tab:Rec_rec}. 

We designed the cohort matrix, such that we can test our model for different scenarios: In the first row, we only altered the diagonal entry, which is indicated in red. In the second row, we changed the diagonal entry and the default entry slightly (indicated in magenta). Last but not least, in the third row we changed the diagonal entry and its neighbours, which is indicated in orange.

The reconstructed rating matrix in \Cref{tab:Rec_rec} has very close values compared to \Cref{tab:Rec_groundtruth}. We tested this with different examples and always saw similar results. One thing to note is that even though we use a Brownian motion as an input for the DNN, the result of the reconstruction was deterministic, so the DNN eliminates the randomness and we leave it for future research to design a DNN with stochastic reconstructions.

\begin{table}[h]
    \caption{Comparison of ground-truth, cohort and reconstructed one-year rating matrix.}
    \addtocounter{table}{-1}
    \label{tab:Rec}

    \begin{subtable}{\textwidth}
        \centering
        \subcaption{Ground-truth rating matrix.}
        \begin{tabular}{|c|*{4}{c}|}
            \hline
            \diagbox{From}{To}      & \rFormat{A} & \rFormat{B} & \rFormat{C} & \rFormat{D} \\
            \hline
            \rFormat{A}     & {\color{red}0.966712} & 0.032916 & 0.000349 & 2.39E-05 \\
            \rFormat{B}     & 0.006511 & {\color{magenta}0.971356} & 0.019862 & {\color{magenta}0.002272} \\
            \rFormat{C}     & 6.64E-06 & {\color{orange}0.008567} & {\color{orange}0.797138} & {\color{orange}0.194288} \\
            \rFormat{D}     & 0     & 0     & 0     & 1 \\
            \hline
        \end{tabular}%
        \label{tab:Rec_groundtruth}
    \end{subtable}
    \bigskip

    \begin{subtable}{\textwidth}
        \centering
        \subcaption{Cohort rating matrix.}
        \begin{tabular}{|c|*{4}{c}|c|}
            \hline
            \diagbox{From}{To}      & \rFormat{A} & \rFormat{B} & \rFormat{C} & \rFormat{D} & $w_t$\\
            \hline
            \rFormat{A}     & {\color{red}0.87004} & 0.032916 & 0.000349 & 2.39E-05 & 0.096671\\
            \rFormat{B}     & 0.006511 & {\color{magenta}0.777085} & 0.019862 & {\color{magenta}0.001817} & 0.194726\\
            \rFormat{C}     & 6.64E-06 & {\color{orange}0.006854} & {\color{orange}0.717424}  & {\color{orange}0.15543} & 0.120285\\
            \rFormat{D}     & 0     & 0     & 0     & 1 & 0\\
            \hline
        \end{tabular}%
        \label{tab:Rec_cohort}
    \end{subtable}
    \bigskip

    \begin{subtable}{\textwidth}
        \centering
        \subcaption{Reconstructed rating matrix.}
        \begin{tabular}{|c|*{4}{c}|}
            \hline
            \diagbox{From}{To}      & \rFormat{A} & \rFormat{B} & \rFormat{C} & \rFormat{D} \\
            \hline
            \rFormat{A}     & 0.964507 & 0.034213 & 0.0006 & 0.000677 \\
            \rFormat{B}     & 0.007204 & 0.966306 & 0.020473 & 0.006023 \\
            \rFormat{C}     & 0.000458 & 0.013529 & 0.802318 & 0.183695 \\
            \rFormat{D}     & 0     & 0     & 0     & 1 \\
            \hline
        \end{tabular}%
        \label{tab:Rec_rec}
    \end{subtable}
\end{table}

\paragraph*{Advantages and limitations.} 
With this new method, we can see that a reconstruction of a cohort rating matrix is possible with larger withdrawal rates and the output of the DNN is close to the ground truth. This is a first step to improve the simple method proposed in \cite{Israel2001}. The choice of a reconstruction method for cohort matrices can have a huge impact for applications and is discussed in more details in \cite{K2022}.

However, for this design of the network, we made the assumption that the operator which is mapping a cohort matrix to an Aalen-Johansen matrix is only additive. In reality, this might not be the case and further studies are needed to include multiplicative changes.

One idea could be to design a new neural network to learn this operator directly by computing cohort rating matrices and Aalen-Johansen rating matrices from the aforementioned historical ratings, maybe with the help of DeepONets (\cite{Lu2019}).

Nevertheless, in a relevant application we will not have access to a ground-truth rating matrix and we will take the uncertainty from this reconstruction in \Cref{sec:LieCalibrationP} into account. 
\subsection{Market Default Probabilities}\label{sec:defaultProb}
As mentioned in the introduction, we will take two sets of data into account for our stochastic rating transition model. Under the historical measure, we will use the cohort rating matrices and its reconstruction by the DNN and under the risk-neutral measure, we have access to the Credit Default Swap (CDS) quotes of individual companies. This means that we can compute market default probabilities of financial sectors, e.g., the corporate sector, by using these CDS quotes.

Usually, the default probabilities under the risk-neutral measure are slightly higher than under the historical measure and we will test our model for three different scenarios of default probabilities, which are shown in \Cref{tab:defaultProbabilities}.

\hyperlink{tab:defaultProbabilitiesOne}{Case 1} is designed to be very close to \Cref{tab:Rec_rec}, \hyperlink{tab:defaultProbabilitiesTwo}{case 2} is more realistic with larger default probabilities for starting rating \rFormat{A} and \rFormat{B}, and \hyperlink{tab:defaultProbabilitiesThree}{case 3} is unrealistic with very high default probabilities to test the model robustness. We will see that our model will perform reasonably well in all scenarios and conjecture that it works equally well in intermediate scenarios.
\begin{table}[h]
    \centering
    \caption{Different scenarios of one-year default probabilities for different starting ratings.}
    \begin{tabular}{*{4}{c}}
        Rating  & \hypertarget{tab:defaultProbabilitiesOne}{Case 1} 
                & \hypertarget{tab:defaultProbabilitiesTwo}{Case 2} 
                & \hypertarget{tab:defaultProbabilitiesThree}{Case 3} \\
        \toprule
        \rFormat{A} & 0.0007 & 0.0068 & 0.0339\\
        \rFormat{B} & 0.0063 & 0.0301 & 0.1506\\
        \rFormat{C} & 0.1929 & 0.1929 & 0.4592\\
        \rFormat{D} & 1 & 1 & 1\\
    \end{tabular}
    \label{tab:defaultProbabilities}
\end{table}
\section{SDEs on the Lie Group of Stochastic Matrices}\label{sec:SDELieGroup}

Our goal for this section is to formulate an SDE that can help to interpolate the generated rating matrices in time. Therefore, we make use of stochastic matrices forming a Lie group and derive an SDE that evolves on this Lie group. Moreover, we show how this SDE can be solved numerically and derive a corresponding theorem of Girsanov for this SDE.

Recall that a Lie group $G$ is a differentiable manifold for which the product is a differentiable mapping $G\times G \to G$. The corresponding Lie algebra $\mathfrak{g}$ to a Lie group $G$ is the tangent space at its identity.

We consider the Lie group of matrices with row sum equal to one, $G =\{R\in\mathrm{GL}(K): R\bm{1}=\bm{1}\}$ with $\bm{1}=[1,\dots,1]^\top\in \R^K$ and its corresponding Lie algebra, $\mathfrak{g}=\{A\in \R^{K\times K}: A\bm{1}=\bm{0}\}$ i.e.\ matrices with row sum zero (see \cite{Coletti2020}). For more details on Lie groups and Lie algebras, we refer the interested reader to \cite{Hall03}.

For the simulation of stochastic matrices we restrict ourselves to a subset of $\mathfrak{g}$, namely
\begin{equation*}
    \mathfrak{g}_{\geq 0} \coloneqq \{A \in\mathfrak{g} : A_{\ell j}\geq 0, \ell\neq j, A_{jj}\leq0, A_{Kj}=0,  \ell,j=1,\dots,K\}.
\end{equation*}
Let $E_{\ell j}$ denote an elementary matrix and define basis matrices $\mathcal{E}_i$ for $i=1,\dots,K-1$ by $E_{\ell j}-E_{\ell\ell}$, more precisely $\mathcal{E}_1=E_{12}-E_{11}$, $\mathcal{E}_2=E_{13}-E_{11}$, $\mathcal{E}_3 = E_{14}-E_{11}$ and so on. 
Then, an arbitrary matrix $A$ in $\mathfrak{g}_{\geq 0}$ can be represented by the linear combination $A=\sum_{i=1}^{(K-1)^2} A^i\mathcal{E}_i$, where we have to assume that $A^i\in\R_{\geq 0}$ for the index $i=1,\dots,(K-1)^2$.

Applying the matrix exponential, $\exp(A)\coloneqq\sum_{k=0}^{\infty}A^k/k!$, to these matrices will give us stochastic matrices $R\in G_{\geq0}$,
\begin{equation}\label{eq:LieGroup}
    G_{\geq 0} \coloneqq\{R\in G: R_{\ell j}\in[0,1], \ell,j=1,\dots,K\},
\end{equation}
\cite[pp.~86\,ff. Chapter 4.2.5: Solving Kolmogorov's Equation]{Stroock2005}, where the last line of matrices $R$ is equal to the last unit vector, which is in accordance with our assumption that the default state is absorbing.
The directional derivative of the matrix exponential along an arbitrary matrix $H$ is given by
\begin{equation*}
    \left(\frac{d}{dA}\exp(A)\right)H = \exp(A)\dD_{-A}(H) \quad \text{with }\; \dD_{-A}(H) = \sum_{k=0}^{\infty}\frac{1}{(k+1)!}\ad_{-A}^k(H),
\end{equation*}
where $\ad_{A}(H)=[A,H]=AH-HA$ denotes the adjoint operator, which is used iteratively,
\begin{equation*}
    \ad_{A}^0(H)=H, \quad \ad_{A}^k(H)=\ad_{A}\big(\ad_{A}^{k-1}(H)\big)=\lbrack A,\ad_{A}^{k-1}(H)\rbrack
\end{equation*}
for $k\geq1$ (see for example \cite[p.~83]{HairerLubichWanner}).
A second order directional derivative of the matrix exponential can be stated via the bilinear operator
\begin{equation*}
    \dDD_{\Sigma}(M,N)=\sum_{n=0}^\infty\sum_{m=0}^\infty\frac{\ad_\Sigma^n(M)}{(n+1)!}\frac{\ad_\Sigma^m(N)}{(m+1)!}+\sum_{n=0}^\infty\sum_{m=0}^\infty\frac{\big[\ad_\Sigma^n(N),\ad_\Sigma^m(M)\big]}{(n+m+2)(n+1)!m!}.
\end{equation*}
A proof can be found in \cite{KPP2021}.

Now, we assume that an SDE for $\generator_t\in \mathfrak{g}_{\geq 0}$ is given by
\begin{equation}\label{eq:SDELieAlgebra}
    d\generator_t = B(t,\generator_t) dt + \sum_{i=1}^{(K-1)^2}C^i(t,\generator_t) dS_t^i,\quad \generator_0=0,
\end{equation}
where $B$ and $C^i$ take values in $\mathfrak{g}_{\geq 0}$ and $S_t^i$ are general semimartingale.
An SDE for $R_t\in G_{\geq 0}$ is obtained by
\begin{multline}\label{eq:SDELieGroup}
		d\rProcess_t =
		\rProcess_t\left(
			\dD_{-\generator_t}\left(B(t,\generator_t)\right)+
			\frac{1}{2}
			\sum_{i=1}^{(K-1)^2}{
				\dDD_{-\generator_t}\left(
					C^i\left(t,\generator_t\right),
					C^i\left(t,\generator_t\right)
				\right)
			}
		\right) dt\\+
		\sum_{i=1}^{(K-1)^2}{
			\rProcess_t\,\dD_{-\generator_t}\left(C^i(t,\generator_t)\right)\,dW_t^{i}, \quad R_0=I
		}.
	\end{multline}
This can be easily verified by applying Itô's lemma to $R_t=R_0\exp(\generator_t)$ in the case $S_t^i=W_t^i$ Brownian motions (as done e.g.\ in \cite{KPP2021}).

\paragraph{Geometric Euler-Maruyama.}
Attempts to solve \eqref{eq:SDELieGroup} numerically for example by using conventional stochastic Runge-Kutta methods might result in a \textit{drift-off}, i.e.\ numerical approximations might leave the manifold and the properties of $R_t$ defined in \eqref{eq:LieGroup} might be violated. 
In order to approximate the solution of \eqref{eq:SDELieGroup} while preserving the Lie group structure of $G_{\geq0}$ on a time interval $[t_k,t_{k+1})$, SDE \eqref{eq:SDELieAlgebra} can be used in the following way:
\begin{enumerate}
    \item Compute an approximation of \eqref{eq:SDELieAlgebra} after one time step by using for example a stochastic Runge-Kutta scheme or an Itô-Taylor expansion and denote it by $\generator_{t_{k+1}}$.
    \item Define a numerical solution of \eqref{eq:SDELieGroup} by $R_{t_{k+1}}=R_{t_k}\exp(\generator_{t_{k+1}})$.
\end{enumerate}
This stochastic Lie group method is based on Runge-Kutta--Munthe-Kaas (RKMK) schemes for ODEs \cite{MuntheKaas1999} and further analysis of stochastic RKMK schemes can for example be found in \cite{LordMalhamSimonWiese,Muniz2021,Muniz2022}.
Note that this method preserves the Chapman-Kolmogorov equation naturally.
Applying the Euler-Maruyama scheme in the first step of this stochastic Lie group method was first called \textit{geometric Euler-Maruyama} in \cite{GoranSolo} and will be the method of choice hereafter. 

Crucial for the numerical solution $R_{t_{k+1}}$ to evolve in $G_{\geq0}$ is the assumption that $\generator_{t_{k+1}}\in\mathfrak{g}_{\geq0}$, which means that \eqref{eq:SDELieAlgebra} must have increasing paths in time.
There are multiple ways to ensure this property, e.g.\ by using jump processes with positive jumps only.
Another possibility involves processes with stochastic coefficients of the form
\begin{align*}
    d\generator_t^i &= a_i(t,Y_t^i) dt, \quad a_i(t,y)\geq 0,\\
    dY_t^i &= b_i(t,Y_t^i)dt + c_i(t,Y_t^i) dS_t, \quad Y_0^i=y_0^i.
\end{align*}
In this case, $\generator_t=\sum_{i=1}^{(K-1)^2} \generator_t^i\mathcal{E}_i$ are positive, pathwise-increasing, continuous stochastic processes for any semimartingale $S_t$.

\paragraph*{Girsanov's theorem.}
For the derivation of the dynamics of our model \eqref{eq:SDELieGroup} under a second measure $\Q$ we apply the standard Girsanov theorem to our SDEs in $\R$. 

	Let the SDE in the Lie algebra satisfy the dynamics under $\P$
	\begin{align*}
		d\generator_t^{\P} &= 
             \sum_{i=1}^{(K-1)^2}{
                \left(
                    b_{i}(t,\generator^{\P,i}_t) dt + 
										c_{i}(t,\generator^{\P,i}_t) dW^{i}_t
                \right) \lieBasis_{i}
             },
	\end{align*}
	where $W^i_t$ are independent standard Brownian motions. Denote $W_t\coloneqq \left(W^1_t,\dots,W^{(K-1)^2}_t\right)$ and let $\mathcal{F}_t$ be the natural filtration of $W_t$. Furthermore, assume that
	the process
	\begin{align*}
		\girsanov_t \coloneqq 
		\exp\left(\int_{0}^{t}{\kappa_s \cdot dW_s}-\frac{1}{2}\int_{0}^{t}{\abs{\kappa_s}^2 ds}\right)
	\end{align*}
	is a $\P$-martingale satisfying $\mathbb{E}^\P\left[\girsanov_t\right]=1$ for adapted, measurable and square integrable processes $\kappa_s \in \R^{(K-1)^2}$. 
	The measure $\Q^\kappa$ given by
	\begin{align*}
		\left.\frac{d\Q^\kappa}{d\P}\right|_{\mathcal{F}_t}\coloneqq \girsanov_t
	\end{align*}
	is well-defined and 
	\begin{align*}
		W^{\kappa}_t \coloneqq W_t - \int_{0}^{t}{\kappa_s ds}
	\end{align*}
	is a Brownian motion under the measure $\Q^\kappa$.
	
	Moreover, assume that the processes $\kappa$ are such that the dynamics of the SDE in the Lie algebra $\lieAlgebra_{\geq 0}$ denoted by $\generator_t^{\kappa}$ are well-defined. 
	Then we have under the new measure 
	\begin{align*}
		d\generator_t^{\kappa} &= 
		 \sum_{i=1}^{(K-1)^2}{
				\left(
						b_{i}(t,\generator^{\kappa,i}_t)dt+ c_{i}(t,\generator^{\kappa,i}_t) \kappa^i_t dt + 
						c_{i}(t,\generator^{\kappa,i}_t) dW^{\kappa,i}_t
				\right) \lieBasis_{i}
		 }\\&\eqqcolon 
		B^\kappa(t,\generator_t^\kappa) dt + \sum_{i=1}^{(K-1)^2}{C^i(t,\generator_t^\kappa) dW_t^{\kappa,i}},
	\end{align*}
	where $B^\kappa(t,\generator_t^\kappa)\coloneqq \sum_{i=1}^{(K-1)^2} \left(b_{i}(t,\generator^{\kappa,i}_t)+ c_{i}(t,\generator^{\kappa,i}_t) \kappa^i_t\right) \lieBasis_{i}$ and
	$C^i(t,\generator_t^\kappa) \coloneqq c_{i}(t,\generator^{\kappa,i}_t) \lieBasis_{i}$.
	
	Moreover, the dynamics of $\rProcess_t$ under the measure $\Q^\kappa$ are given by
	\begin{align*}
		d\rProcess_t^\kappa &=
		\rProcess_t^\kappa\left(
			\dD_{-\generator_t^\kappa}\left(B^\kappa(t,\generator_t^\kappa)\right)+
			\frac{1}{2}
			\sum_{i=1}^{(K-1)^2}{
				\dDD_{-\generator_t^\kappa}\left(
					C^i\left(t,\generator_t^\kappa\right),
					C^i\left(t,\generator_t^\kappa\right)
				\right)
			}
		\right) dt\\&\quad+
		\sum_{i=1}^{(K-1)^2}{
			\rProcess_t^\kappa\,\dD_{-\generator_t^\kappa}\left(C^i(t,\generator_t^\kappa)\right)\,dW_t^{\kappa,i}
		}.
	\end{align*}
One sufficient condition for this change of measure to be valid is the positivity of 
$\generator_t^{\kappa,i}$ $\P$-almost surely for all $t\geq 0$ and $i=1,\dots,(K-1)^2$. 

We will see in \Cref{sec:LieCalibrationQ} how to apply this for the calibration to the market default probabilities.

\section{Numerical Experiments}\label{sec:numerics}
In this section, we will use the geometric Euler-Maruyama approach to approximate an SDE on the Lie group of stochastic matrices. We calibrate the resulting rating model $\rSDE_t$ at $t=1$, i.e.\ one year, under the historical measure to the reconstructed rating matrix in \Cref{tab:Rec_rec} and explain how to account for uncertainity coming from this reconstruction. This is described in \Cref{sec:LieCalibrationP} in more details. In \Cref{sec:LieCalibrationQ}, we will compare two popular methods for the change of measure and calibrate it to the default probabilities in \Cref{tab:defaultProbabilities}.
In both sections, we will discuss the distributions of the individual entries in the rating transition matrices and study some properties rating matrices should satisfy. These properties are introduced next in \Cref{sec:errors}. 

For the calibration of the rating SDE we used \matlab with the \matlabGOtoolbox
and for the training of the TimeGAN and the reconstruction DNN \python with \tensorflow
running on \OS, on a machine with the following specifications: processor
\CPU and \RAM, and a \GPU.

\subsection{Rating Properties}\label{sec:errors}
We introduced the following properties in \cite{KM2022}
to estimate the quality rating matrices up to one year: 
\begin{compactenum}
    \item It is more likely to stay in the initial rating than changing to another: This means rating matrices are strongly diagonally dominant, i.e.\ for $i=1,\dots,K$
        \begin{align}
            \left[R_{t}\left(\omega\right)\right]_{ii} \geq 
            \sum_{j \neq i}{\left[R_{t}\left(\omega\right)\right]_{ij}}.
            \label{eq:sDD}
        \end{align}
    \item Downgrading is more likely than upgrading: This means that the sum of the upper triangular matrix     is bigger than the sum of the lower triangular matrix, i.e.
        \begin{align}
            \sum_{i<j}{\left[R_{t}\left(\omega\right)\right]_{ij}}\geq 
            \sum_{i>j}{\left[R_{t}\left(\omega\right)\right]_{ij}}.
            \label{eq:dML}
        \end{align}
    \item Lower rated entities are more likely to default: This means that the default column is increasing from best starting rating to lowest, i.e. 
        \begin{align}
            \left[R_{t}\left(\omega\right)\right]_{1K}\leq 
            \left[R_{t}\left(\omega\right)\right]_{2K}\leq\dots\leq
            \left[R_{t}\left(\omega\right)\right]_{KK}.
            \label{eq:mDC}
        \end{align}
    \item The rating spreads more over time: We measure this by looking for decreasing diagonal elements, i.e.\ for all $s<t$ and all  $i=1,\dots,K$
        \begin{align}
            \left[R_{s}\left(\omega\right)\right]_{ii}\geq \left[R_{t}\left(\omega\right)\right]_{ii}.
            \label{eq:iRS}
        \end{align}
\end{compactenum}
These properties are not strict in the sense that they can be violated on some occasions. Moreover, one might think of other properties for rating matrices. Also, for long term rating matrices (more than 1 year) these properties might not hold true anymore. This makes it very hard to define rigorous conditions for  rating matrices in general and is subject to future research and economical validation.

These properties were perfectly satisfied for the reconstructed rating matrices $\rRec_t$. 

For our stochastic model, we will discuss these properties after the individual calibrations under the historical and risk-neutral measure in more details.

\subsection{Calibration under the Historical Measure}\label{sec:LieCalibrationP}
In this subsection, we will explain how we can calibrate $\rGEM_t$ to $\rRec_t$, while taking reconstruction uncertainty into account. Therefore, let us first of all explain what we understand as reconstruction uncertainty.

In an application of the proposed reconstruction via the DNN in \Cref{sec:ratingCohort}, there is no way to know how close this reconstruction would have been to the rating matrix without withdrawals. This introduces uncertainty in the method. We will present a method, which highlights the key factors for extracting this uncertainty from the reconstruction and the cohort matrix but caution the reader that these choices are arbitrary and discuss in \Cref{sec:Conclusion} how we would like to change this approach in the future.

We will assume that $\rRec_t$, $t=1$ year, will be the mean of our stochastic model, i.e., we set
\begin{gather*}
    \left[\mu_{1}^{\mathrm{gEM}}(t)\right]_{ij} \coloneqq \frac{1}{M} \sum_{w=1}^{M}  \left[\rGEM_t(w)\right]_{ij},\\
    f_1 \colon\Pi \rightarrow \R^{K^2},\quad
         f_1(p)\coloneqq \vectorize\left(\mu_{1}^{\mathrm{gEM}}(t;p) - \rRec_t\right),
\end{gather*}
where $p\in \Pi$ denote the parameters of our model, and $M\in \N$ is the number of simulations.

To take uncertainty into account, we would like to extract volatility information from $\rCohort_t$
and $\rRec_t$. Therefore, let us first of all have a look in \Cref{tab:distanceRC}.
\begin{table}[h]
    \centering
    \caption{$L^1$-distance of $\rCohort_t$ (\Cref{tab:Rec_cohort}) and $\rRec_t$ (\Cref{tab:Rec_rec}).}
    \begin{tabular}{|c|*{4}{c}|}
        \hline
        \diagbox{From}{To}      & \rFormat{A} & \rFormat{B} & \rFormat{C} & \rFormat{D} \\
        \hline
         \rFormat{A} & {\color{red}0.0945} & 0.0013 & 2.5123e-04 & 6.5335e-04\\ 
         \rFormat{B} & 6.9289e-04 & {\color{magenta}0.1892} & 6.1092e-04 & {\color{magenta}0.0042}\\ 
         \rFormat{C} & 4.5115e-04 & {\color{orange}0.0067} & {\color{orange}0.0849} & {\color{orange}0.0283}\\ 
         \rFormat{D} & 0 & 0 & 0 & 0\\
         \hline
    \end{tabular}
    \label{tab:distanceRC}
\end{table}
This serves as an indicator of how much the DNN had to \enquote{repair} the cohort matrix $\rCohort_t$ in each individual entry. We observe that this distance matrix, which we will denote by $D_t$, mirrors the changes we made to the ground-truth matrix. In the first row, the diagonal has the lion's share of the needed reparation and in the second row we recognize that the default column was changed as well and the proportions in the third row match as well. 

However, the distance alone is not a good estimator of uncertainty because it does not take the geometrical structure of the stochastic matrices into account. Let us explain this in the first row. We know that our rating model will have row sums equal to one, meaning that a huge volatility in the diagonal cannot be compensated by very small volatilities in the other components. Thus, the uncertainty in the diagonal element is actually very small. In the second row, we see a similar situation. In the third row, this is more spread out. To measure this spread, we take the row sums of the distance matrix $D_t$, which we denote by $\eta_t$, and compute the proportions of the entire reconstruction distance row-wise denoted by $\rho_t$, i.e., for $i,j=1,\dots,K$ we define
\begin{align*}
    \left(D_t\right)_{ij} &\coloneqq 
        \abs{\left(\rRec_t-\rCohort_t\right)_{ij}},\\
    \eta_t &\coloneqq 
        \left(\sum_{j}^{K} \left(D_t\right)_{ij} \right)_{i=1,\dots,K},\\
    \left(\rho_t\right)_{ij} &\coloneqq 
        \frac{
            \left(D_t\right)_{ij}
        }{ 
            \left(\eta_t\right)_{i}
        }\1_{\left(\eta_t\right)_{i} > 0},\\ 
    \rAdjusted_t&\coloneqq 
        \rCohort_t + \rho_t \odot D_t,
\end{align*}
where $\odot$ denotes the elementwise or Hadamard product.

The idea is to adjust the cohort method by equalizing the initial distances $D_t$. If there is a disproportionate large distance in one entry compared to the others, then the majority of the distance for each row of $D_t$ will be added to $\rCohort_t$. If they are similar, then a similar distance will be added.

\begin{table}[h]
    \centering
    \caption{Values of $\rAdjusted_t$ using $\rCohort_t$ (\Cref{tab:Rec_cohort}) and $\rRec_t$ (\Cref{tab:Rec_rec}).}
    \begin{tabular}{|c|*{4}{c}|}
        \hline
        \diagbox{From}{To}      & \rFormat{A} & \rFormat{B} & \rFormat{C} & \rFormat{D} \\
        \hline
         \rFormat{A} & 0.9624 & 0.0329 & 3.4924e-04 & 2.8275e-05\\ 
         \rFormat{B} & 0.0065 & 0.9610 & 0.0199 & 0.0019\\ 
         \rFormat{C} & 8.3277e-06 & 0.0072 & 0.7773 & 0.1621\\ 
         \rFormat{D} & 0 & 0 & 0 & 1\\
         \hline
    \end{tabular}
    \label{tab:distanceRC}
\end{table}

In \Cref{tab:distanceRC}, we show the adjusted matrix $\rAdjusted_t$ using $\rCohort_t$ (\Cref{tab:Rec_cohort}) and $\rRec_t$ (\Cref{tab:Rec_rec}).
We can see that the diagonal elements in the first and second row are now a lot closer to $\rRec_t$ and the overall distances of $\rRec_t$ and $\rAdjusted_t$ are similar. Now, it makes sense to take a two-point variance of $\rRec_t$ and $\rAdjusted_t$ to include uncertainty, i.e.
\begin{gather*}
\left[\mu_{2}^{\mathrm{gEM}}(t)\right]_{ij} \coloneqq 
    \frac{1}{M-1} \sum_{w=1}^{M}   
        \left(
            \left[\rGEM_t(w)\right]_{ij}-
            \left[\mu_{1}^{\mathrm{gEM}}(t)\right]_{ij}
        \right)^2,\\
    f_2:\Pi\rightarrow \R^{K^2}, \quad
        f_2(p)\coloneqq 
            \vectorize\left(\mu_{2}^{\mathrm{gEM}}(t;p) - (\rRec_t-\rAdjusted_t)^2\right),
\end{gather*}
where all operations on the matrices are understood elementwise.

In total, we define our objective function as
\begin{align*}
    f(p)\coloneqq \left[w_1\cdot f_1(p),w_2\cdot f_2(p)\right]^\top,
\end{align*}
where $w_1,w_2 \in \R_{>0}$ are weights, which we set to one in our experiments.

The minimisation problem can now be formulated as a non-linear least square problem
\begin{align}
    \min_{p\in \Pi} \norm{f(p)}_2^2.
    \label{eq:nonLinearLeastSquare}
\end{align}

As aforementioned, the reader may use a different method to account for reconstruction uncertainty and this serves as an illustration which properties to consider.

Let us now assume that each of the SDEs are given by 
\begin{align*}
    dA_t^i &= \abs{Y_t^i}^{a_i} dt,\\
    dY_t^i &= b_i dt + \sigma_i dW_t,\quad Y_0^i=0.
\end{align*}
They have a parameter for a constant drift $b_i$, power $a_i$ and volatility $\sigma_i$, which are all assumed to be positive.
The parameter set is therefore given by positive real numbers 
$\Pi^{\mathrm{gEM}}\coloneqq \R^{3\cdot (K-1)^2}_{\geq 0}$ by stacking the individual parameters below each other.
We found during our experiments that values between zero and three worked best. 
We calibrated $\rGEM_t$ for $t=1$, i.e.\ for the 12 month rating transitions, with the procedure from above.  The corresponding parameters after the calibration with $M=1000$ trajectories for $\rGEM_t$ can be found in \Cref{tab:gemParamCal1}. The first column explains to which basis element the coefficients belong to. To be more precise, $2-3$ means starting rating is 2 and at $t=1$ we transition to rating 3. The minimisation error \eqref{eq:nonLinearLeastSquare} in this case was
$2.046e-07$, resulting in an excellent fit and it took roughly 497 seconds using \lsqnonlin with the Levenberg-Marquardt algorithm.

\begin{figure}[h]
    \centering
    \includegraphics[width=\columnwidth]{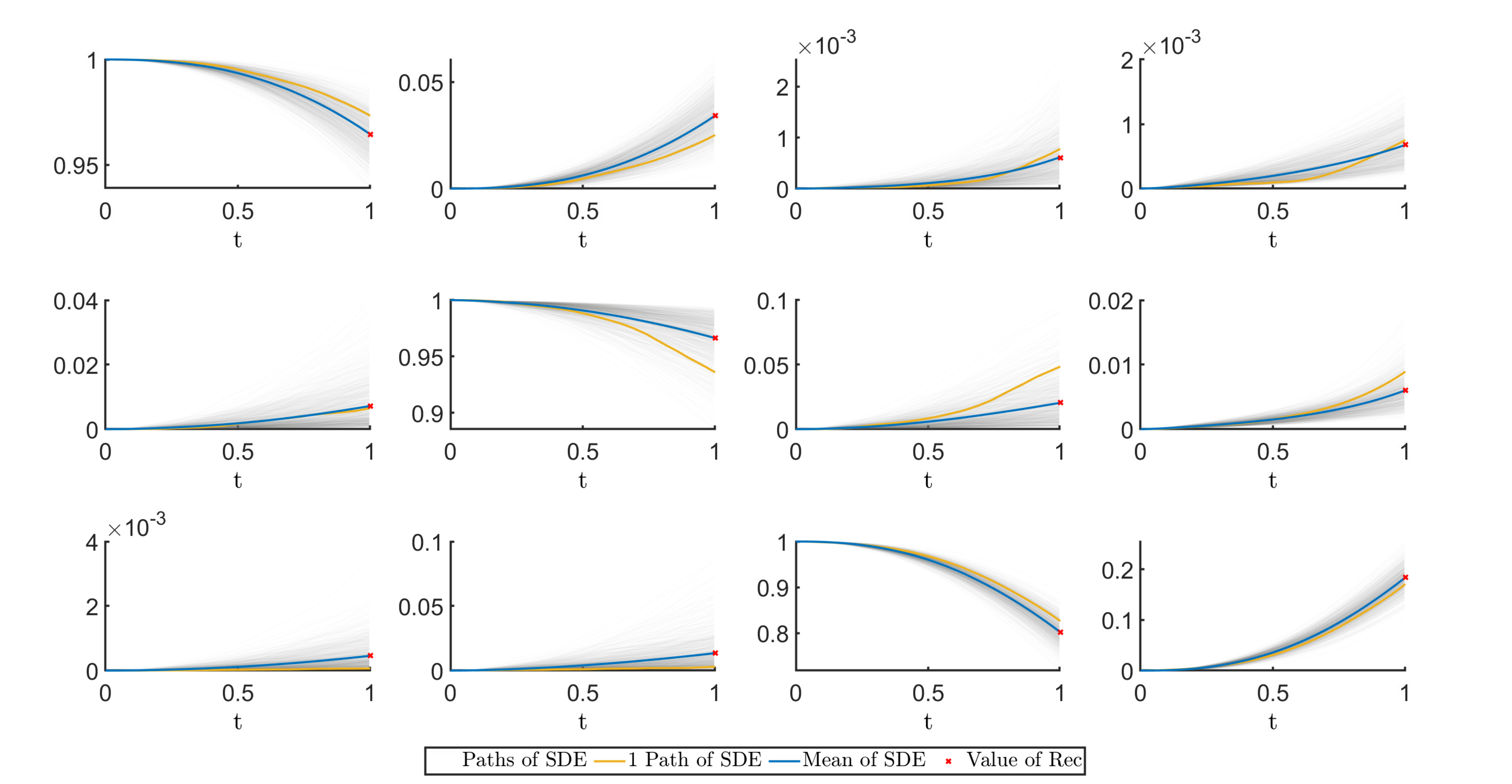}
    \caption{Trajectories of calibrated $\rGEM_t$ with parameters as in \Cref{tab:gemParamCal1}.}
    \label{fig:gemTra1}
\end{figure}

\begin{table}[htbp]
    \centering
    \caption{Parameters of $\rGEM$ after calibration at $t=1$.}
    \begin{tabular}{*{4}{c}}
        From-To & $a$ & $b$ & $\sigma$\\
        \toprule
        1-2 & 1.57e+00 & 4.69e-02 & 5.76e-02\\
        1-3 & 1.62e+00 & 8.83e-03 & 1.15e-02\\
        1-4 & 1.53e+00 & 2.05e-01 & 3.22e-02\\
        2-1 & 1.54e+00 & 6.91e-02 & 9.68e-02\\
        2-3 & 1.44e+00 & 3.14e-03 & 5.75e-03\\
        2-4 & 1.46e+00 & 9.98e-02 & 9.50e-02\\
        3-1 & 7.87e-01 & 1.42e-04 & 1.10e-04\\
        3-2 & 5.76e-01 & 1.00e-04 & 1.00e-04\\
        3-4 & 1.51e+00 & 6.42e-01 & 5.10e-02
    \end{tabular}
    \label{tab:gemParamCal1}
\end{table}
In \Cref{fig:gemTra1}, we can see the trajectories of $\rGEM_t$ over time for each entry in the rating matrix except for the last row. The upper left corner are the transition probabilities from \rFormat{A}
to \rFormat{A}, right next to it from \rFormat{A} to \rFormat{B} and so on. The grey lines are a cloud of
$1000$ trajectories of $\rGEM_t$ and the yellow line is one trajectory. The blue line is the mean at each time of the process and the red dots are the values of $\rRec_t$ at $t=1$ year.
We can see a good fit at the terminal time to $\rRec_t$ by comparing how close the mean of $\rGEM_t$ is compared to the values of $\rRec_t$. 
\paragraph*{Analysis of the rating distributions and properties}

\begin{figure}[h]
    \centering
    \includegraphics[width=\columnwidth]{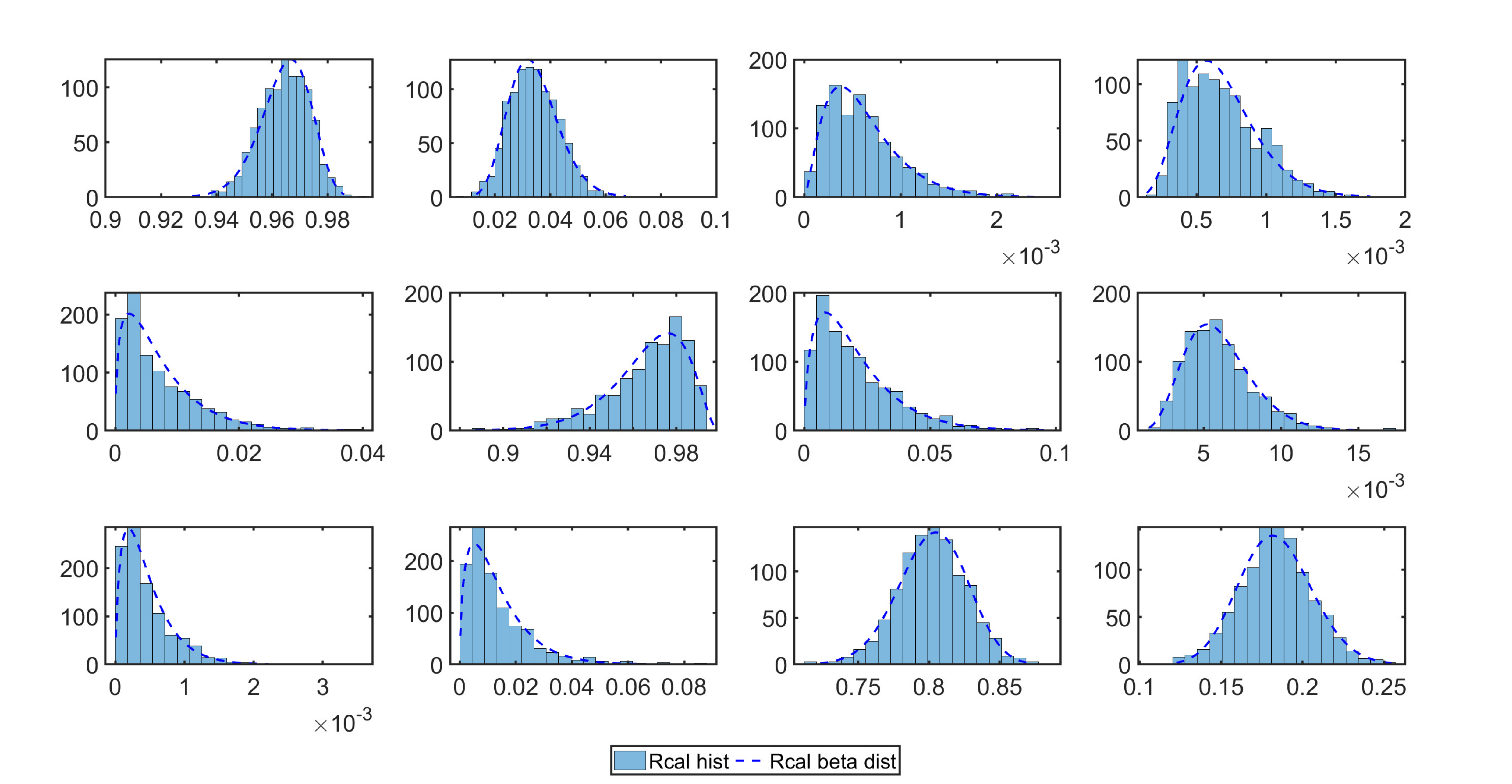}
    \caption{Histograms of ratings transition probabilities at 12 months.}
    \label{fig:gemRD4}
\end{figure}

Let us focus for the moment on \Cref{fig:gemRD4}, i.e., the rating transitions for one year. 
In the upper left corner, we can see the transition probabilities from \rFormat{A} to \rFormat{A}, right next to it \rFormat{A} to \rFormat{B}, and so on.
The blue columns are the histograms of transition probabilities after one year in the calibrated $\rGEM_t$ model and the dark blue dashed line is a fitted beta distribution to the histograms.
We see a close match of the beta distribution to the histograms. 

Moreover, we can see in this picture the impact of our uncertainty approach very clearly. In the last row, we can see a spread of 10\,\% in the diagonal entry, compared to a 4\,\% spread in the diagonal entry of the first row, which is the desired behaviour. This mitigates the impact of the reconstruction method for the cohort rating matrix.
Let us now discuss the conditions \eqref{eq:sDD}--\eqref{eq:iRS} for this model. From $t=0.25$ onwards all of the properties were perfectly satisfied. Till $t=\frac{1}{12}$ roughly $3.7\,\%$ of the trajectories violated the monotone default column property between rating \rFormat{B} and \rFormat{C} by a few basis points, which is the overall worst violation. Including time-dependent parameters could remove this small violation completely. 
\subsection{Calibration under the Risk-Neutral Measure}\label{sec:LieCalibrationQ}
In this subsection, we will show how to change the measure of
\begin{align*}
    d\generator_t^i &= \abs{Y_t^i}^{a_i} dt,\\
    dY_t^i &= b_i dt + \sigma_i dW^i_t,\quad Y_0^i=0,
\end{align*}
and calibrate the Girsanov kernel $\kappa$ in such a way to the market default probabilities that the mean of the model matches the market data.

Let us assume that the processes $\kappa_t\equiv \kappa \in \R^{(K-1)^2}$ are constant, then the dynamics of $\generator_t^i$ are given by
\begin{align*}
    d\generator_t^{\kappa,i} &= \abs{Y_t^{\kappa,i}}^{a_i} dt,\\
    dY_t^{\kappa,i} &= \left(b_i + \sigma_i \kappa_i\right) dt + \sigma_i dW^{\kappa,i}_t.
\end{align*}
In this case, no further conditions on $\kappa$ are required to ensure a valid change of measure such that $\generator_t^\kappa \in \lieAlgebra_{\geq 0}$. The process $\rProcess^{\mathrm{gEM},\kappa}_t$ can again be computed by the geometric Euler method.

As we have seen in \Cref{sec:defaultProb}, we only have access to the default probabilities under a risk-neutral measure. Therefore, we can only expect to calibrate as many parameters of the change of measure as we have default data; in this case three parameters for \rFormat{A}, \rFormat{B}, and \rFormat{C}.

In the literature, there are two popular choices how to define $\kappa$ using only $K-1$ parameters. One was introduced by Jarrow, Lando and Turnbull (cf. \cite{Jarrow1997}), which we will refer to as the JLT change of measure, by multiplying each row of the rating matrix (except the default row), with the same parameter. The other change of measure is called \emph{exponential change of measure} used by \cite{Bielecki2012} in the context of rating modelling and goes back to \cite{Palmowski2002}. For this change of measure, a certain ratio of the $K-1$ parameters is used, which we will now explain in detail.

Therefore, let us define
\begin{align*}
	D \coloneqq \sum_{k=1}^{(K-1)^2}{\kappa_k \lieBasis_k} \in \R^{K\times K}.
\end{align*}
and denote the entries of $D$ by $d_{ij}$, $i,j=1,\dots,K$. Let us consider a vector $h\in \R^{K}$, such that $h_K=1$, i.e., we have $K-1$ free parameters for our change of measure.

Now, for the JLT change of measure each row of $D$ has the same parameter $\kappa_k$. To make this more precise, set $\kappa_k$, $k=1,\dots, (K-1)^2$, such that $d_{ij}=h_i$ for all $j=1,\dots,K$, $i=1,\dots,K-1$ with $i\neq j$.

Similarly, for the exponential change of measure we set $\kappa_k$, $k=1,\dots, (K-1)^2$, such that
$d_{ij}=\frac{h_i}{h_j}$ for all $j=1,\dots,K$, $i=1,\dots,K-1$ with $i\neq j$ and furthermore assume  $h_i\neq 0$.

For the calibration under $\Q$, we assume that $a_i,b_i,\sigma_i$ are already known from the calibration under the historical measure as seen in \Cref{sec:LieCalibrationP}. Therefore, we are looking for the values of $h$, such that the mean of the default probabilities of our model, i.e., the last column of $\rProcess^{\mathrm{gEM},h}_t$, are close to the market default probabilities $\PD(t)$. We will consider here for simplicity the case, where we calibrate at the terminal time $T=1$ year.

To be more precise, we are looking for a solution to the minimization problem
\begin{align}
	\min_{h\in \R^{K-1}\setminus \left\{0\right\}} \norm{\rProcess^{\mathrm{gEM},h}_T \, e_K - \PD(T)}_2^2.
	\label{eq:nonLinearLeastSquareQ}
\end{align}
As explained in \Cref{sec:defaultProb}, we are considering the three different cases of the market default probabilities in \Cref{tab:defaultProbabilities} using the JLT and exponential change of measure.

The computational times for the calibration using a \lsqnonlin with the Trust-Region-Reflective algorithm were in all cases ranging from $25$ to $360$ seconds for the JLT change of measure and ranging from $50$ to $90$ seconds for the exponential change of measure.

\begin{table}%
\caption{Parameters of $\rGEM$ after calibration at $t = 1$ to market default probabilities using the coefficients \Cref{tab:gemParamCal1}.}
\centering
\begin{tabularx}{\linewidth}{l*{2}{cC}}
$\PD(T)$ & 
	\multicolumn{2}{c}{Exp. change of measure} & 
	\multicolumn{2}{c}{JLT change of measure}\\
\toprule
 & error \eqref{eq:nonLinearLeastSquareQ} & h
 & error \eqref{eq:nonLinearLeastSquareQ} & h\\
\midrule
\hyperlink{tab:defaultProbabilitiesOne}{Case 1} & 
	1.057e-06 &	$\left[0.33,\,  0.26,\,  0.66 \right]$ & 
	4.539e-16 & $\left[0.12,\,  0.07,\,  0.49  \right]$\\
\hyperlink{tab:defaultProbabilitiesTwo}{Case 2} & 
	2.569e-12 &	$\left[12.38,\,  3.07,\,  0.46 \right]$ & 
	8.310e-08 & $\left[15.22,\,  6.20,\,  0.47 \right]$\\
\hyperlink{tab:defaultProbabilitiesThree}{Case 3} & 
	2.780e-06 &	$\left[198.29,\,  151.86,\,  13.73\right]$ & 
	1.060e-03 & $\left[21.76,\,  15.13,\,  31.66\right]$\\
\end{tabularx}
\label{tab:gemParamCal2}
\end{table}

In \Cref{tab:gemParamCal2}, we can see the results of the calibration \eqref{eq:nonLinearLeastSquareQ}. The errors were all excellent except for the JLT change of measure in the case of unrealistically high default probabilities.  
We can also see the calibrated parameters $h$ corresponding to the different changes of measure. We did not print the last value of $h_K$, since it is equal to one.

\begin{figure}%
	\includegraphics[width=\columnwidth]{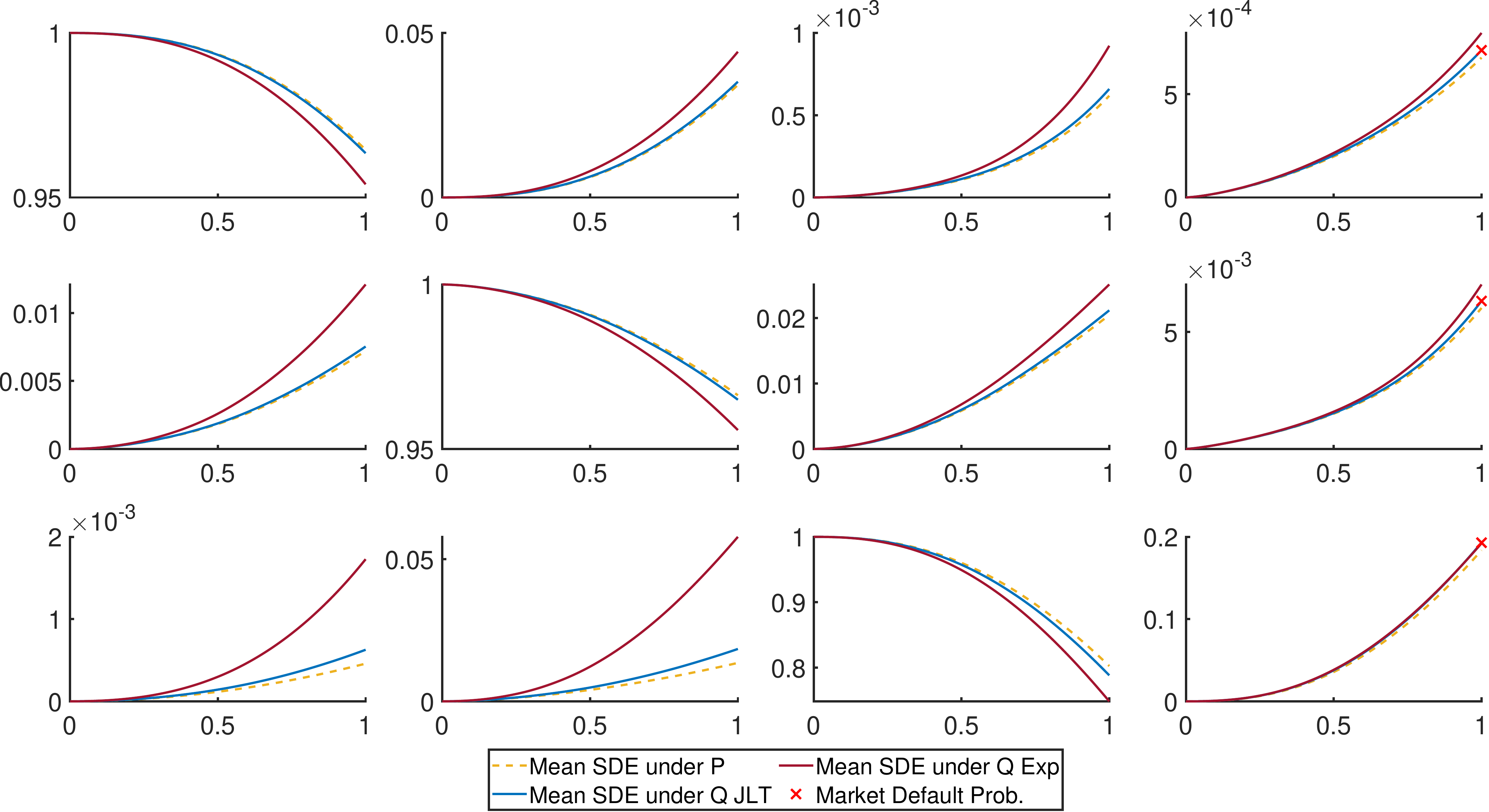}%
	\caption{Comparison of JLT and exponential change of measure of the mean over all trajectories of $\rGEM_t$ using case 1 of the market default probabilities in \Cref{tab:defaultProbabilities}.}%
	\label{fig:gemCOM1}%
\end{figure}

\begin{figure}%
	\includegraphics[width=\columnwidth]{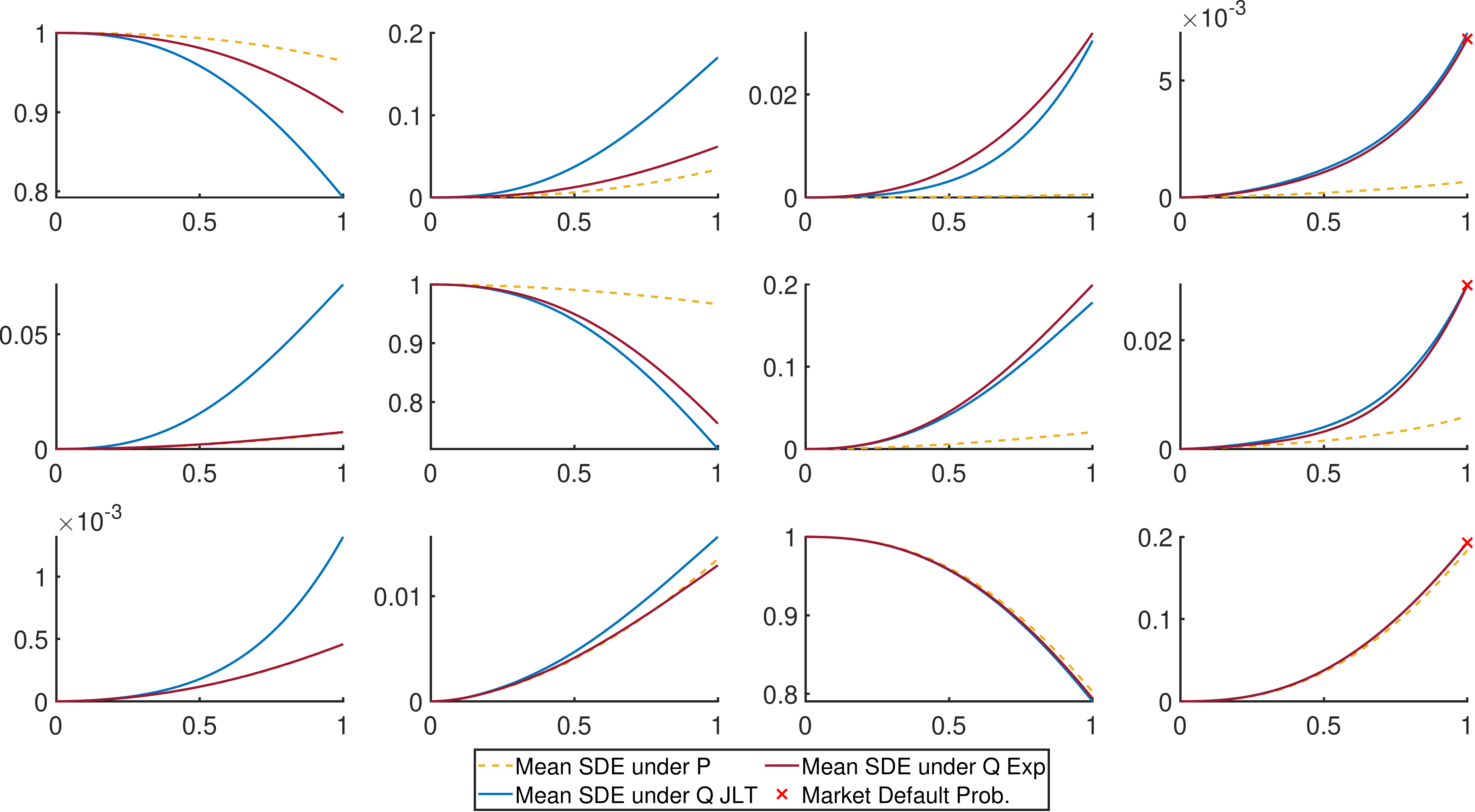}%
	\caption{Comparison of JLT and exponential change of measure of the mean over all trajectories of $\rGEM_t$ using case 2 of the market default probabilities in \Cref{tab:defaultProbabilities}.}%
	\label{fig:gemCOM2}%
\end{figure}

\begin{figure}%
	\includegraphics[width=\columnwidth]{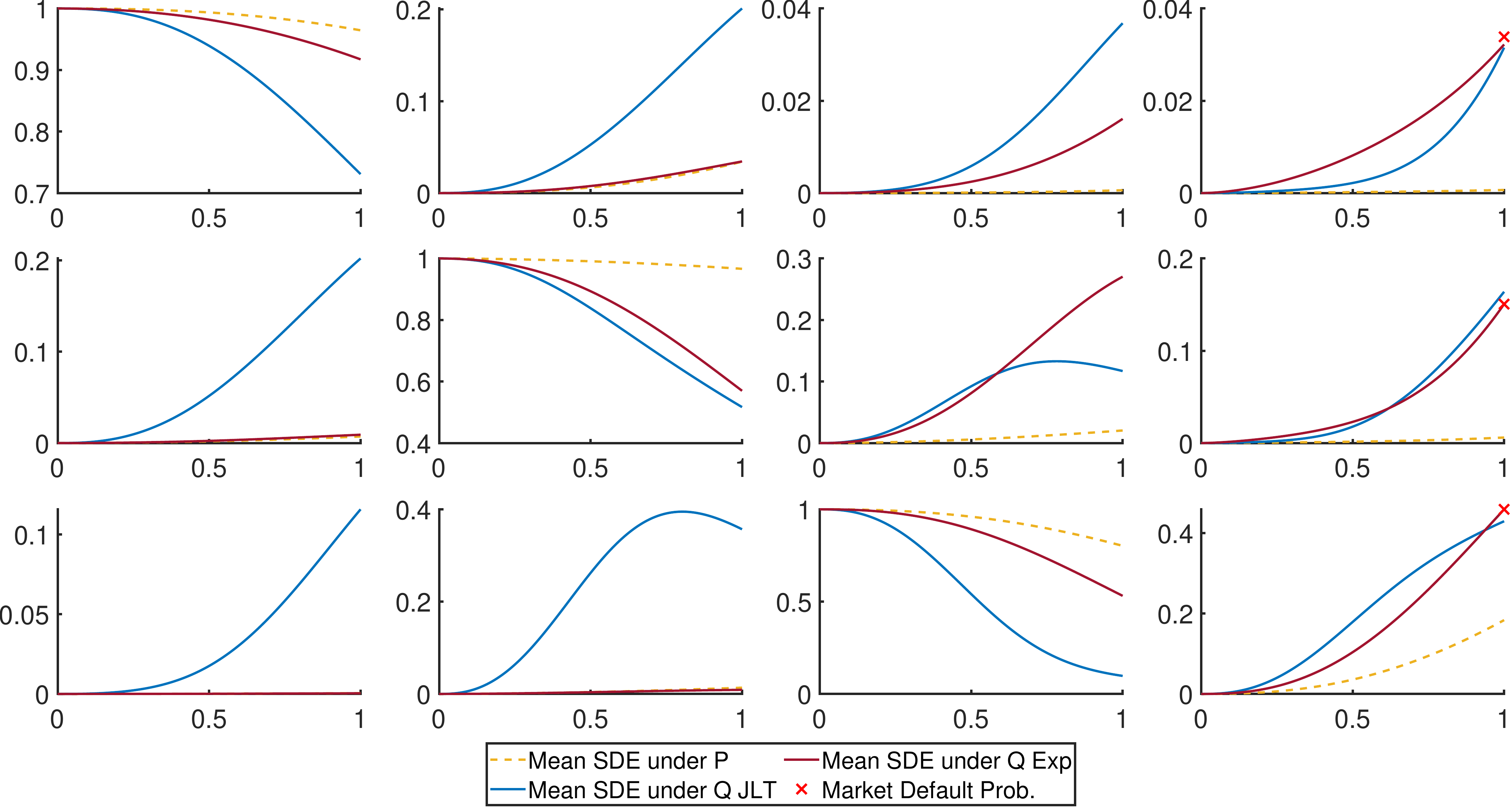}%
	\caption{Comparison of JLT and exponential change of measure of the mean over all trajectories of $\rGEM_t$ using case 3 of the market default probabilities in \Cref{tab:defaultProbabilities}.}%
	\label{fig:gemCOM3}%
\end{figure}

Since the structure of the change of measure differs a lot from each other, we can expect a very different impact on the rest of the rating transition matrix, which we can see in \Cref{fig:gemCOM1}--\ref{fig:gemCOM3}. We show the evolution of the mean over $M=1000$ trajectories from $t=0$ up to $t=1$ year corresponding to the market default probabilities in \Cref{tab:defaultProbabilities}. The yellow dashed line corresponds to the mean under the historical measure to compare the impact of the individual change of measure. The blue bold line depicts the JLT change of measure and the dark red bold line the exponential change of measure. The bright red crosses are the market default probabilities at $t=1$. Each figure is divided into a matrix of subfigures corresponding to the entries in the rating matrix, i.e.\ the upper left corner describes the transition probabilities from \rFormat{A} to \rFormat{A} at each time $t$ and the one right next to it from \rFormat{A} to \rFormat{B}, and so on. The last row of the rating matrix is excluded, since its constant.

In \Cref{fig:gemCOM1}, where the default probabilities under the risk-neutral measure are close to the ones under the historical measure, we see that both changes of measure impact the entire rating matrix very similar and the JLT change of measure is closer to the rating transitions under $\P$. This changes for 
\Cref{fig:gemCOM2} and \Cref{fig:gemCOM3}. The JLT change of measure overestimates the impact on the rating matrix apart from the default column and the exponential change of measure is closer to the transitions under the historical measure. Let us focus on the subfigure for the transition from \rFormat{C} to \rFormat{C} in \Cref{fig:gemCOM3} using \hyperlink{tab:defaultProbabilitiesThree}{case 3}. We can see that the JLT change of measure breaks down, it predicts that almost no company would stay at the rating \rFormat{C} after one year, which is totally unrealistic. On the other hand, the exponential change of measure performs very well even in this case.

Therefore, we suggest to use the exponential change of measure.


\paragraph*{Analysis of the rating distributions and properties.}
Let us now have a closer look at the distributions and rating properties for the exponential change of measure.

\begin{figure}%
	\centering
	\includegraphics[width=\columnwidth]{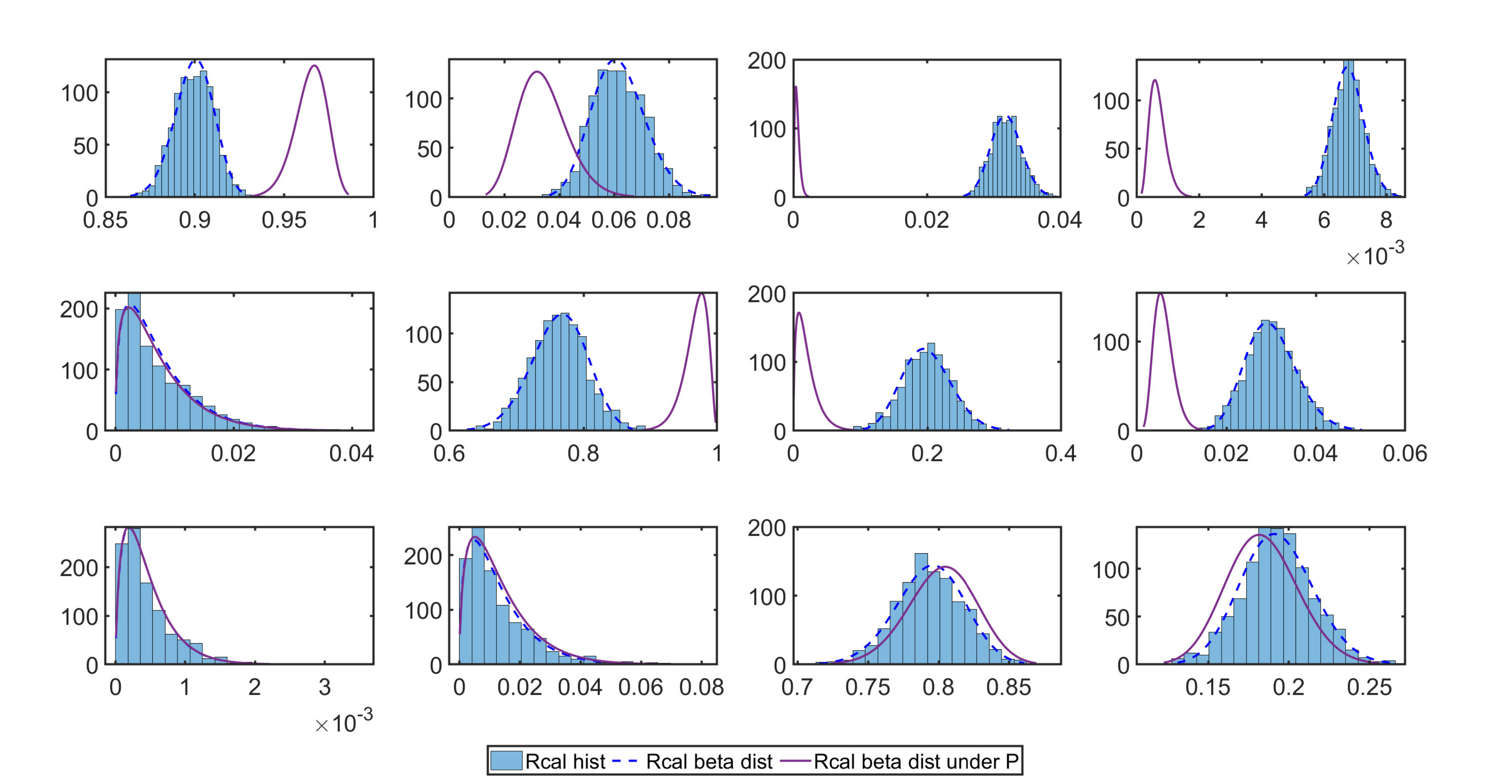}%
	\caption{Histograms of rating transition probabilities at 12 months of $\rGEM_t$ using case 2 of the market default probabilities in \Cref{tab:defaultProbabilities}.}%
	\label{fig:gemRDQ1}%
\end{figure}

For the rating distributions, we consider the second case of market default probabilities in \Cref{tab:defaultProbabilities}, which are slightly more elevated compared to the historical default probabilities. The other cases show similar results. In \Cref{fig:gemRDQ1}, the histogram of the distribution of $\rGEM_t$ under measure $\Q^\kappa$ corresponds to the light blue columns and a fitted beta distribution to this histogram is illustrated by the dark blue dashed line. In purple, we show the fitted beta distribution under the measure $\P$.  This figure contains again subfigures corresponding to the entries in the rating matrix after one year, where the last row is again excluded.
We can see that the change of measure impacts both, the mean and the spread of the distribution under the new measure. The larger the difference between the risk-neutral and historical default probabilities, the greater the impact. This can be seen by comparing the first and second row to the third row. For the third row we kept a close default probability under both measures and the distributions both look similar. 

In the first and second row, we see the greatest difference for the downgradings and the diagonal elements. The other distributions for the upgradings are almost identical to the historical transitions, which matches the observation we made in 
\Cref{fig:gemCOM1}--\ref{fig:gemCOM3}.

Now, let us briefly summarize the rating properties \eqref{eq:sDD}--\eqref{eq:iRS}. We can judge whether these properties are satisfied or not from \Cref{fig:gemCOM1}--\ref{fig:gemCOM3} and refer for the details to the \matlab code.

For both changes of measure, we see similar results under the historical measure in \hyperlink{tab:defaultProbabilitiesOne}{case 1} and \hyperlink{tab:defaultProbabilitiesOne}{case 2}. Almost all the properties were perfectly satisfied with a few minor exceptions. For the unrealistically high default probabilities in \hyperlink{tab:defaultProbabilitiesOne}{case 3}, we see bigger violations. For example, at $t=0.25$ years $35\,\%$ of the trajectories violate the monotone default column property between \rFormat{B} and \rFormat{C} using the exponential change of measure. For the JLT change of measure, roughly $30\,\%$ violate the property that downgradings are more likely at $t=1,6$ months and even $40\,\%$ at $t=3$ months. 

These huge violations are due to the unrealistic scenario and demonstrate that the rating properties serve as a good instrument to judge whether the rating model is working or not. In the realistic scenarios they are satisfied, while they are violated the most in an unrealistic scenario.

\section{Nested Stochastic Simulation Algorithm}\label{sec:LieSimulation}
In this section, we show how to sample a rating process $\ratingProcess_t$ from the rating transition model $\rGEM_t$. For this purpose, we propose the following method: First, let us fix a path of the stochastic rating transition model $\rGEM_t\left(\omega\right)$. Then, we use the definition of the geometric Euler scheme for this path, i.e.
\begin{align*}
    \rGEM_{t_{k+1}}\left(\omega\right) = \rGEM_{t_k}\left(\omega\right) \exp\left(\generator_{t_{k+1}}\left(\omega\right)\right), 
\end{align*}
where $\generator_{t_{k+1}}$ are the one-step approximations of our SDE in $\mathfrak{g}_{\geq0}$. Now, on each interval $[t_{k},t_{k+1})$ we will approximate $\generator_{t_{k+1}}$ by
\begin{align*}
    \generator_{t_{k+1}}\left(\omega\right) \overset{!}{=} \tilde{\generator}_{k+1} \left(t_{k+1}-t_{k}\right),
\end{align*}
where $\tilde{\generator}_{k+1}\in \lieAlgebra_{\geq 0}$ is a constant matrix for $k=0,\dots,N$ and $N\in\N$ is the number of points in our time grid. We will assume that the time grid is homogeneous and denote its mesh size by $\Delta t$. Rearranging the previous equations leads to
\begin{align*}
    \tilde{\generator}_{k+1} = \frac{1}{\Delta t} \generator_{t_{k+1}}\left(\omega\right).
\end{align*}
With the help of this approximation, we define a piecewise generator of an inhomogeneous continuous-time Markov chain (ICTMC)
\begin{align*}
    \tilde{\generator}_t \coloneqq \sum_{k=0}^{N-1}{\tilde{\generator}_{k+1} \1_{[t_k,t_{k+1})}(t)}.
\end{align*}
To shorten the notation we will refer to this piecewise ICTMC as PHCTMC.

There are several techniques in the literature concerning the efficient simulation of ICTMCs and its forward equation. For a detailed discussion, we refer to 
\cite{Li2012} and \cite{Arns2010} among many others. We will recall the procedure described in \cite{K2022} for this special case of PHCTMCs:

We will use the Gillespie Stochastic Simulation Algorithm (SSA) (cf. \cite{Gillespie2007}), which is also called \emph{Kinetic Monte Carlo (KME)} method, on each sub-interval $[t_{k},t_{k+1})$, where the PHCTMC is homogeneous. It turns out that this approach is very fast, because our state space has only a few states. Let us recall that the path $\omega$ from the beginning is still fixed and for one path of $\rGEM_t$ the simulation works as follows:

The PHCTMC is homogeneous on each interval
$[t_{k-1},t_k]$, $k=1,\dots,n$, by construction and now we iterate over those intervals, i.e., assume that we are already at $t=t_{k-1}$ with current rating $i$. On each subinterval we proceed as follows: 
\begin{compactenum}[(i)]
	\item If $t\leq t_k$ and $\left(\tilde{\generator}_k\right)_{ii} \neq 0$ draw two uniform random numbers $r_1$, $r_2$, otherwise end and set $\ratingProcess^{i_0}_t=i$ on $[t,t_k]$;
	\item Retrieve the exponentially distributed transition waiting time with parameter
		$-\left(\tilde{\generator}_k\right)_{ii}$ as
		\begin{align*}
			\tau = 
			\frac{-\log\left(r_1\right)}{-\left(\tilde{\generator}_k\right)_{ii}}=
			\frac{\log(r_1)}{\left(\tilde{\generator}_k\right)_{ii}}.
		\end{align*}
		If $t+\tau\geq t_k$ set $\ratingProcess^{i_0}=i$ and go to the next interval, starting with step (i), else continue to calculate the next state;
	\item Now, sample from the discrete state transition distribution
		$\left[\frac{\left(\tilde{\generator}_k\right)_{ij}}{-\left(\tilde{\generator}_k\right)_{ii}}\right]_{j\neq i}$. This can be done by choosing the first integer $j$, such that
		$\sum_{k=1,k\neq i}^{j}{\frac{\left(\tilde{\generator}_k\right)_{ij}}{-\left(\tilde{\generator}_k\right)_{ii}}}> r_2$, which is equivalent to
		\begin{align*}
			\min_j
			\sum_{l=1,l\neq i}^{j}{\left(\tilde{\generator}_k\right)_{il}}>-\left(\tilde{\generator}_k\right)_{ii}r_2.
		\end{align*}
		Now, go back to (i) with $\ratingProcess^{i_0}_{t+\tau}=j$.
\end{compactenum}
Like in a nested Monte-Carlo simulation for doubly stochastic processes, we will first sample $M_1\in\N$ generators $\rGEM_t$ and simulate conditioned on each path $M_2\in\N$ trajectories of the rating model with the aforementioned algorithm. In total, we will have $M\coloneqq M_1 \cdot M_2 \in\N$ paths.

\begin{table}%
\caption{Simulation errors of $\rGEM_t$ using case 2 of \Cref{tab:defaultProbabilities} as market default probabilities. First row is the mean error of the model rating transitions and simulated rating transitions under the historical measure. The second row contains the errors under the risk-neutral measure of the model and simulated rating transitions.}
\centering
\begin{tabular}{|c|*{4}{c}|} 
	\hline
	\diagbox{Error}{Time}&$t=\frac{1}{12}$ & $t=\frac{3}{12}$ & $t=\frac{6}{12}$ & $t=1$\\
	\hline
	$\frac{1}{K^2}\norm{R_t^{\P} - R_t^{\text{Sim},\P}}_{\R^{K,K}}$ & 
	7.7545e-04 & 0.0017 & 0.0030 & 0.0052\\
	$\frac{1}{K^2}\norm{R_t^{\Q} - R_t^{\text{Sim},\Q}}_{\R^{K,K}}$ &
	8.3978e-04 & 0.0023 & 0.0044 & 0.0083\\
	\hline
\end{tabular}
\label{tab:GEMsimulationErrors}
\end{table}

We calculated the \emph{nested SSA} for each initial rating in parallel on a CPU and sampled 
$M=M_1\cdot M_2$ with $M_1=1000$ different rating transition operators simulated by $\rGEM_t$ and 
$M_2=10000$ SSA
trajectories. This took roughly 64 minutes using a machine with $64$ CPU cores. 
As a side note, sampling only $100\cdot 100=10000$ paths takes roughly 16 seconds with 6 CPU cores.

Let us now judge the accuracy of this simulation technique in this setting. Therefore, let us have a look at \Cref{tab:GEMsimulationErrors}.
The errors were computed by first calculating the transition matrices from the simulated rating processes $R^{i,P}_t$, $P=\P,\Q$, by counting how many trajectories are at each state and dividing by the total amount of trajectories. The result of this is denoted by 
$R^{\text{Sim},P}_t$ and we used the Frobenius norm divided by the squared number of ratings and the mean over all trajectories to evaluate the error. We can see that the errors are still satisfactorily small.

 In our experiments, we found that for the inner simulations for the SSA $M_2$ should be at least $1000$ to guarantee a small simulation error.

\begin{figure}
\centering
\includegraphics[width=.5\columnwidth]{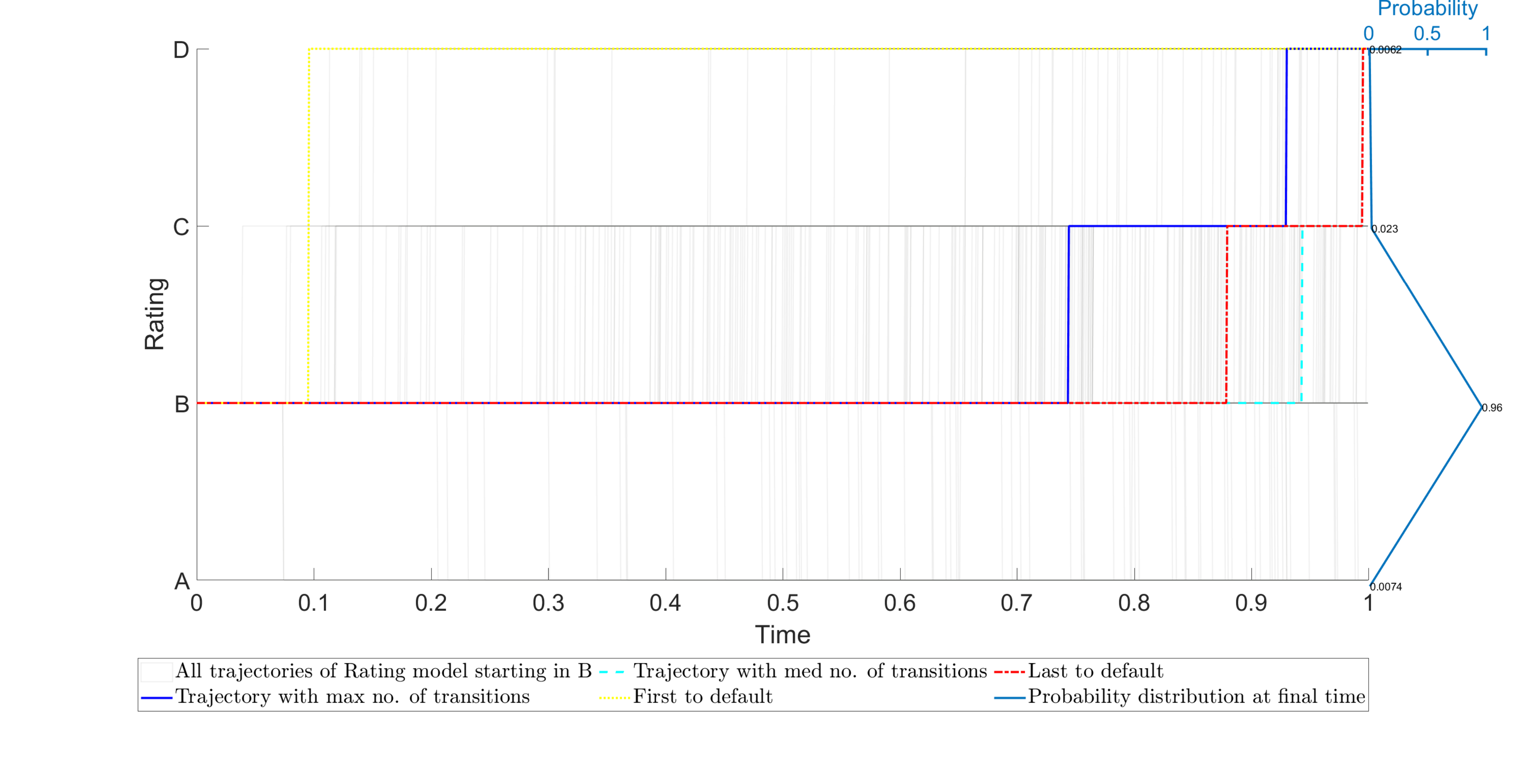}
\includegraphics[width=.5\columnwidth]{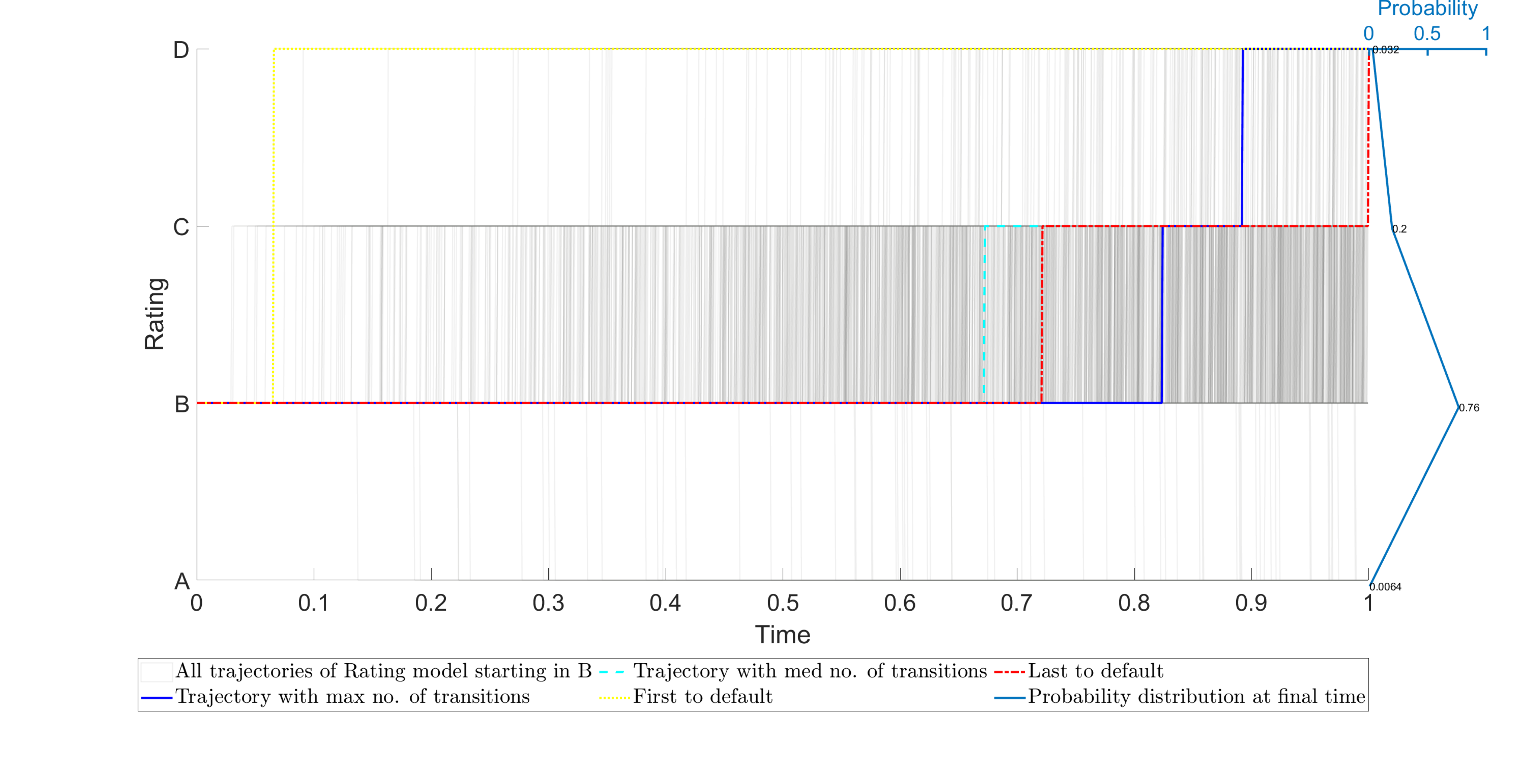}
\caption{Simulated trajectories $X_t$ using the transition operators from $\rGEM_t$ starting in rating \rFormat{B}. The left picture is under measure $\P$ and the right picture is under measure $\Q$ using case 2 of \Cref{tab:defaultProbabilities} as market default probabilities.}%
\label{fig:GEMsimulation1}%
\end{figure}

In \Cref{fig:GEMsimulation1}, we can see an example of the simulated ratings $\ratingProcess_t$ using $\rGEM_t$ under the historical measure in the top figure and under the risk-neutral measure in the bottom picture. We used the exponential change of measure with the mild probabilities of default (case 2) in \Cref{tab:defaultProbabilities}. The grey lines illustrate $M=100\cdot 100=10000$ different paths of $\ratingProcess_t$ and the highlighted paths in different colors are some particular examples of trajectories. We notice that the transitions under the risk-neutral measure compared to the historical ones, reveal a huge difference. We can see from the deep grey areas in the bottom picture that a lot more transitions occurred under the risk-neutral measure. Consistently with \Cref{fig:gemCOM2}, we can see the probability distribution of the ratings at $t=1$ year on the right-hand side. 
The number of transitions seems to increase the more we progress in time in a smooth fashion, which we consider to be realistic.

\section{Application to Rating Triggers for Collateral-Inclusive Bilateral
Valuation Adjustments}\label{sec:ratingTriggers}
In this section, we study bilateral credit and debit valuation adjustments (hereafter referred to as \textCCVA and \textCDVA) of a portfolio of trades between two parties having signed a collateral agreement dependent on ratings. 
The main references for the general theory of XVA are 
\cite[\ pp.~375\,ff. Chapter 12.3 Credit Valuation Adjustment and Risk Management]{Oosterlee2019}
and
\cite[\ pp.~305\,ff. Chapter 13 Collateral, Netting, Close-Out and Re-Hypothecation]{Brigo2013}.
Recall that \textCCVA and \textCDVA are adjustments to the fair price of a financial portfolio accounting for the potential loss in case of default of the counterparty and the owner, respectively. These are usually defined as an average of the exposure (positive and negative, respectively) weighted by the probability of default. The two parties usually sign a so-called netting agreement, so as to consider the exposure at portfolio level (as opposed to trade-wise). Attached to the netting agreement, one often has a Credit Support Annex (CSA) by which each of the two parties further agrees to interchange securities (referred to as collateral) to reduce the exposure of the other party. In the case of bankruptcy, the collateral account can be used to mitigate the losses of the non-defaulting party, although collateral is often non-segregated and therefore also at risk. Since posting collateral is another expense for an entity, it is desirable to keep the postings as small as possible while simultaneously keeping the losses due to a default event small as well. To achieve this goal, more and more CSAs specify thresholds of permitted unsecured exposure in terms of the credit quality of the parties: the higher the credit quality of a party, the smaller the amount collateral it has to post (and the larger the unsecured exposure of the other party). 

A customary way to measure the credit quality of an entity is to use credit ratings. A high rating means that the entity is very likely to fulfill its financial obligations towards its contracting party, while a low rating associates an increasing risk for meeting the aforementioned obligations. In this line of thought, the default can be viewed as the worst possible rating. CSAs dependent on ratings are said to have rating triggers: a change of rating of one of the parties triggers a change of threshold of that party.

Since the exposure depends on the amount of collateral posted or received, to compute \textCCVA and \textCDVA in presence of a CSA with rating triggers, it is necessary to model the rating processes of the contracting parties. 

In this section, we will use $\rGEM_t$ as our rating transition model and refer to \cite{K2022} for similar tests using an inhomogeneous continuous-time Markov chain model.

Due to the doubly-stochastic rating process $X_t$ corresponding to $\rGEM_t$ being pathwise Markovian, we would like to point out that the memoryless property of such processes makes them somewhat unrealistic as models for the rating process, since an entity with a history of successive downgrades is more likely to be considered risky than a competitor with long-time constant rating. 

 
In most of the literature, time-homogeneous models for rating transitions are considered. However, it is empirically evident that a time-homogeneity assumption is often violated. We have already seen in our previous experiments that we can calibrate our fully inhomogeneous model in a meaningful way to the data, which is an improvement to the existing models in the literature.

For a more detailed discussion of the Markovianity and time-homogeneity, we refer the reader to \cite{Lencastre2014}.

Throughout this section, we are taking the point of view of a bank having a portfolio of deals with a counterparty. The two parties have signed a netting set agreement with a CSA having rating-dependent thresholds. The stochastic process representing the future mark-to-market of the portfolio is denoted by $V_t$. We will assume that both contracting parties are subject to default and the default time will be denoted by $\tau_{B}$ and $\tau_{C}$ for the bank and the counterparty, respectively.
Additionally, we will suppose that the same rating matrices apply to both, i.e.\ they are in the same industrial sector and the consideration of two different sectors is discussed in \Cref{rem:twoSectors}.
For our illustration, we will assume that the bank has the highest rating today, whereas the counterparty has a mid-range rating today, since we expect a bank to default less likely than the majority of companies. The evolution of their ratings over time will be denoted by $\ratingProcess_t^B$ and $\ratingProcess_t^C$, respectively, and we will set $\ratingProcess_t \coloneqq \left(\ratingProcess_t^B,\ratingProcess_t^C\right)$ to shorten the notation.

Let $C_t$ be the stochastic process representing the value of the collateral account. In particular, $C_t>0$ if the collateral is received by the bank. We will assume for simplicity that $C_t$ depends on $V$ only through the value $V_t$: more precisely we will suppose that $C_t \coloneqq f\left(V_t,\ratingProcess_t\right)$. In particular to avoid path dependencies, we assume there are no minimum transfer amounts, the impact of this assumption being not material for our purposes.

We will discuss the following three scenarios of collateral agreements:

\begin{compactenum}
\item \emph{uncollateralized}, i.e.\ no collateral is interchanged and $f\equiv 0$;
\item \emph{perfectly collateralized}, i.e.\ collateral is posted instantaneously  at a discrete set of times, e.g.\ daily, and is equal to the mark-to-market ($V_t = C_t$), $f(v,r) = v$;
\item \emph{rating-trigger dependent}, more precisely we will focus on the case of thresholds depending on ratings
(see below for a description of $f$ in this case).
\end{compactenum}

The relation of the bank to the counterparty is illustrated in Figure \ref{fig:bankCpty}
and reads as follows\footnote{
We will use the same conventions as in \cite[\ pp.~310\,ff. Chapter 13.2 Bilateral CVA Formula under Collateralization]{Brigo2013}, in particular $X^+=\max\left(X,0\right)$ and 
$X^-=\min\left(X,0\right)$.
}:

To illustrate the impact of collateral in risk mitigation, let us assume instantaneous posting and no rehypothecation of collateral for simplicity. Assume the counterparty (but not the bank) defaults at time $\tau$. Then on the one hand we have the value of the portfolio $V_\tau$ and on the other hand we have the value of the collateral account $C_{\tau}$. 
We distinguish four cases:
\begin{itemize}
\item $V_\tau\geq 0, C_{\tau}\geq 0$: The portfolio generates a positive exposure for the bank but this is mitigated by the collateral (which can be fully retrieved by the bank because of no rehypotecation). Therefore, the outstanding claim is $V_\tau - C_{\tau}$. 
\item $V_\tau\geq 0, C_{\tau} \leq 0$: Although the portfolio generates a positive exposure for the bank, the bank had posted collateral just before default. Because of no rehypothecation, the bank can fully get back its collateral and the outstanding claim is therefore $V_\tau$.
\item  $V_\tau\leq 0, C_{\tau} \geq 0$: The counterparty gets back the collateral posted to the bank and also gets the value of the portfolio $\abs{V_\tau}$.
\item $V_\tau\leq 0, C_{\tau} \leq 0$: The counterparty keeps the collateral posted by the bank and also gets the remaining value of the portfolio $\abs{V_\tau-C_{\tau}}$.
\end{itemize}
The behaviour in case of default of the bank is symmetrical.

Additionally, the individual collateral postings depending on the collateral agreement are depicted by $f\left(V_t,\ratingProcess_t\right)^-$, meaning that the bank has to post collateral if this value is greater than zero and its analogue for the counterparty is given by $f\left(V_t,\ratingProcess_t\right)^+$.
 
A comprehensive explanation of all default events in this bilateral setup can be found in 
\cite[pp.~311--312 Chapter 13.2.1 Collection of CVA Contributions]{Brigo2013}.

\tikzset{mnode/.style={draw,rectangle,black,align=center,
											 text width = 2.5cm}}
\tikzset{marrow/.style={draw,thick}}
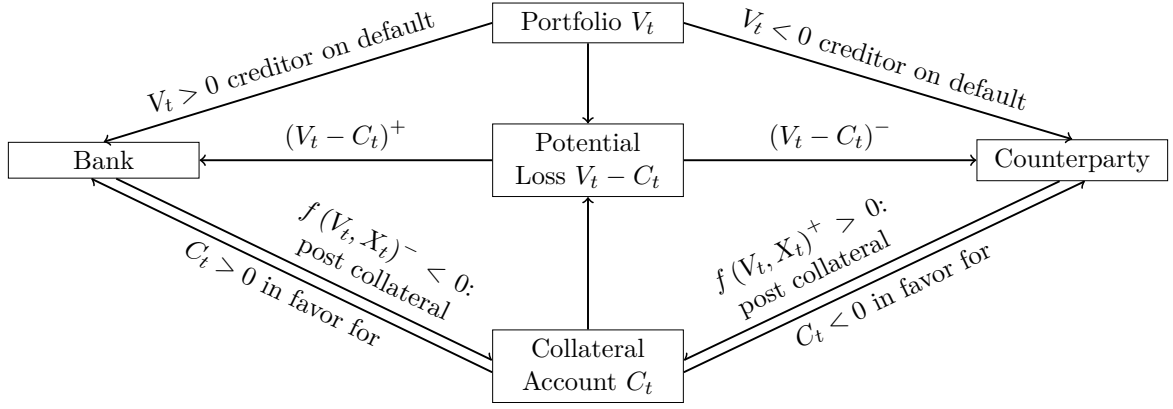
\begin{figure}%
\begin{minipage}[c][][c]{\linewidth}
\centering
\resizebox{\linewidth}{!}{
\begin{tikzpicture}
	\node[mnode] (bank) at (-7,0) {Bank};
	\node[mnode] (cpty) at (7,0) {Counterparty};
	\node[mnode] (portfolio) at (0,2) {Portfolio $V_t$};
	\node[mnode] (collateral) at (0,-3) {Collateral Account $C_t$};
	\node[mnode] (exposure) at (0,0) {Potential Loss $V_t-C_t$};
	\draw[->,marrow] (portfolio) -- (exposure);
	\draw[->,marrow] (collateral) -- (exposure);
	\draw[->,marrow] (exposure) -- node[midway,above] {$\left(V_t-C_t\right)^+$} (bank);
	\draw[->,marrow] (exposure) -- node[midway,above] {$\left(V_t-C_t\right)^-$} (cpty);
	
	\path[name path=bb] (bank.south west) -- (bank.south east);
	\path[name path=colb] ([yshift=2.5pt]collateral.west)  -- ([yshift=2.5pt]bank.south);
	\path[name intersections={of=bb and colb, by={ColB}}];
	\draw[-left to,marrow] 
		(ColB)
			-- node[text width= 3.75cm,midway,above,sloped,xshift=2.5em,align=center] {$f\left(V_t,\ratingProcess_t\right)^-<0$: post collateral} 
		([yshift=2.5pt] collateral.west) ;
	
	\path[name path=bcol] ([yshift=-2.5pt]collateral.west) -- ($([yshift=-2.5pt]collateral.west)!10cm!([yshift=-2.5pt]bank.south)$);
	\path [name intersections={of=bb and bcol, by={BCol}}];
	\draw[-left to,marrow] 
		([yshift=-2.5pt]collateral.west) 
			to node[midway,below,sloped] {$C_t>0$ in favor for} 
		(BCol);
	
	\path[name path=cc] (cpty.south west) -- (cpty.south east);
	\path[name path=cptyc] ([yshift=2.5pt]collateral.east)  -- ([yshift=2.5pt]cpty.south);
	\path[name intersections={of=cc and cptyc, by={CptyC}}];
	\draw[-right to,marrow] 
		(CptyC)
			-- node[text width= 3.75cm,midway,above,sloped,xshift=-2em,align=center] {$f\left(V_t,\ratingProcess_t\right)^+>0$: post collateral} 
		([yshift=2.5pt] collateral.east) ;
	
	\path[name path=ccpty] ([yshift=-2.5pt]collateral.east) -- ($([yshift=-2.5pt]collateral.east)!10cm!([yshift=-2.5pt]cpty.south)$);
	\path [name intersections={of=cc and ccpty, by={CCpty}}];
	\draw[-right to,marrow] 
		([yshift=-2.5pt]collateral.east) 
			to node[midway,below,sloped] {$C_t<0$ in favor for} 
		(CCpty);
	\draw[->,marrow] 
		(portfolio.west) 
			to node[midway,above,sloped] {$V_t>0$ creditor on default} 
		(bank.north);
	\draw[->,marrow] 
		(portfolio.east) 
			to node[midway,above,sloped] {$V_t<0$ creditor on default} 
		(cpty.north);
		
	\pgfresetboundingbox
	\path[use as bounding box] let \p{coordsBank} = (bank.west),
						\p{coordsCpty} = (cpty.east),
						\p{coordsC} = (collateral.south),
						\p{coordsV} = (portfolio.north)
				in
				($(\x{coordsBank},\y{coordsC})$) rectangle 
				($(\x{coordsCpty},\y{coordsV})$);
\end{tikzpicture}
}
\end{minipage}
\caption{Illustration of bank and counterparty relations in terms of exposure and collateral agreements.}%
\label{fig:bankCpty}%
\end{figure}

Next, we will discuss the impact of rating triggers compared to the aforementioned scenarios of collateral agreements on collateral-inclusive CVA, DVA and BVA, followed by a discussion on the pre-default distribution of the rating processes in 
Section \ref{sec:preDefault}.
\subsection{XVA with Different Collateral Agreements}\label{sec:CXVA}
We are interested in the impact of rating triggers on \textCBVA, \textCCVA and \textCDVA (Collateralized Bilateral, Credit, Debit Valuation Adjustments) without the possibility of rehypothecation and zero interest rate at mid-market to simplify the investigation.

Before we dive into this topic, let us first of all discuss our benchmark portfolio.
Since our main purpose is to analyse the general behaviour of the rating model, we are not interested in setting up an accurate model for the computation of $V_t$. In particular, the $V_t$ that we consider does not represent the value of a portfolio of real deals. Rather we decided to simulate $V_t$ using a number of independent Brownian motions with different volatilities and life-times $l^i$ to account for the cash-flows of the portfolio. To be more precise
\begin{align*}
	V_t \coloneqq 
	V_0+\sigma_0 W^0_t+\sum_{i=1}^{n}{
		\sigma_i W^i_t \1_{t\leq l^i}
	},
\end{align*}
where $V_0 \in \R_{\geq 0}$ is the initial value, $W^i$, $i=0,\dots,n$, $n\in\N$, are independent Brownian motions, $\sigma_i \in \R$ are volatilities and $l^i \in [0,T]$ are uniformly distributed random variables describing the different life-times of the cash-flows.
In the experiment we use $V_0=0$, $n=24$ and $\sigma_i$ are the standard normal random variables multiplied by $10$ for scaling and its sign indicates a positive or negative cash-flow (from the bank perspective). Also, notice that we designed the portfolio in such a way that at least one cash-flow survives till $T$ by not adding a finite life-time to $W_t^0$. 
 
 Now, let us briefly recall the relevant definitions of \textCXVA\newline (cf. \cite[\ p.~314 Equation 13.4, \ p.~316 Equation 13.10]{Brigo2013}) without re-hypothecation
\begin{align}
	\CBVA\left(t,T,\collateral\right)&\coloneqq
	\CDVA\left(t,T,\collateral\right)-
	\CCVA\left(t,T,\collateral\right),
	\label{eq:CBVA}\\
	\CDVA\left(t,T,\collateral\right)&\coloneqq
	-\mathbb{E}^{\Q}\left[
		\left.
		\1_{\tau=\tau_B<T}\,
		\lgd_B\,
		\left(V_\tau^- - C_\tau^-\right)^-
		\right|
		\mathcal{G}_t
	\right],
	\label{eq:CDVA}\\
	\CCVA\left(t,T,\collateral\right)&\coloneqq
	\mathbb{E}^{\Q}\left[
		\left.
		\1_{\tau=\tau_C<T}\, 
		\lgd_C \,
		\left(V_\tau^+ - C_\tau^+\right)^+
		\right|
		\mathcal{G}_t
	\right],
	\label{eq:CCVA}
\end{align}
where $\mathcal{G}_t$ is the filtration containing all the default-free market information plus default monitoring. These values are calculated under a risk-neutral measure $\Q$, which explains why we were interested in changing the measure of our rating model from the historical probabilities $\P$ to the risk-neutral measure in the first place.

The evaluation of the collateral account at the exact time of the default event, i.e. $C_\tau$, might seem confusing. We could imagine a scenario in which bonds or stocks could be used as collateral, making it necessary to evaluate the collateral account at the default event. In our case, we will assume that the collateral account will be a pure cash account, meaning that upon a default event the value will not be updated from its previous value $C_{\tau-}$. Therefore, it is very important to study the distribution of ratings prior to default, which is subject to Section \ref{sec:preDefault}.

We now describe the function $f\left(\exposure_{t},\ratingProcess_{t}\right)$ in the case of rating-triggers dependent agreements, following \cite[\ pp.~316\,ff. Chapter 13.5.2 Collateralization Through Margining]{Brigo2013}.
Let $r_i^{x}\geq 0$, $x\in \left\{B,C\right\}$, $i=1,\dots,K$ denote the threshold for the party $x$ in case $x$ has rating $i$: this means that the maximum unsecured exposure of the other party will be at most $r_i^{x}$. 

Now, we introduce the rating triggers $\rho^x$ with corresponding thresholds $r_i^x$ as
\begin{align*}
	\rho^x(i)\coloneqq
	\sum_{j=1}^{K}{r_j^x \1_{j}(i)}.
\end{align*}
As a small example, setting for all $i=1,\dots, K$ and $x=\left\{B,C\right\}$ the thresholds $r_i^x=+\infty$ lead to the uncollateralized scenario and $r_i^x=0$ to the perfectly collateralized scenario.

The amount of collateral to be posted by the bank at time $t_j$ is then
\begin{align*}
	\left(\exposure_{t_j} + \rho^B(\ratingProcess^B_{t_j})\right)^- -C_{t_j-}^-.
\end{align*}
For the counterparty we have analogously
\begin{align*}
	\left(\exposure_{t_j} - \rho^C(\ratingProcess^C_{t_j})\right)^+-C_{t_j-}^+.
\end{align*}

As aforementioned, we assume for simplicity that the value $C_{t_j}$ of the collateral account at time $t_j$ is equal to $C_{\beta(t_j)}$ where $\beta(u)$ is the last collateral posting date before $u$. In particular, we assume there is no remuneration on the collateral account.
We then have
\begin{align*}
	C_{t_0}\coloneqq 0,\quad C_{t_n}\coloneqq 0, \quad C_{u-}\coloneqq
	C_{\beta(u)}.
\end{align*}
\begin{align*}
	C_{t_j}\coloneqq
	C_{t_j-}+
	\left(\left(\exposure_{t_j} + \rho^B(\ratingProcess^B_{t_j})\right)^- -C_{t_j-}^-\right)+
	\left(\left(\exposure_{t_j} - \rho^C(\ratingProcess^C_{t_j})\right)^+-C_{t_j-}^+\right).
\end{align*}
This can be rewritten as
\begin{align*}
	C_{t_j}=
		\left(\exposure_{t_j} + \rho^B(\ratingProcess^B_{t_j})\right)^-+
		\left(\exposure_{t_j} - \rho^C(\ratingProcess^C_{t_j})\right)^+
		\eqqcolon
		f\left(\exposure_{t_j},\ratingProcess_{t_j}\right).
\end{align*}

\begin{figure}%
\begin{tikzpicture}[spy using outlines={rectangle, connect spies}]
	\node[draw=none] at (0,0) {
		\includegraphics[width=\columnwidth]{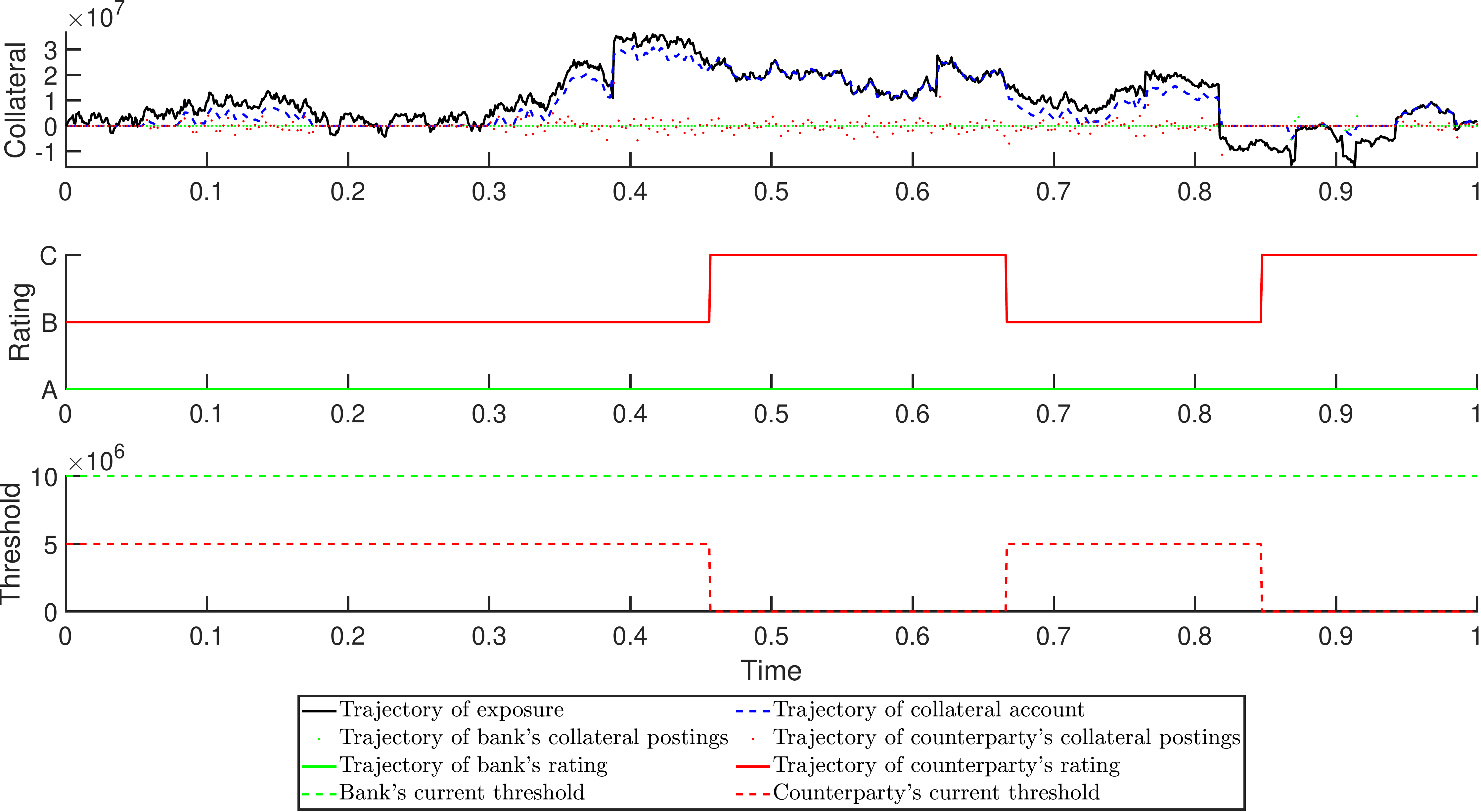}
	};
	
	\spy [orange, draw, height = 1cm, width = 2cm, magnification = 2] on (-6.6,2.9) in node at (-6,4.8);
	\node[anchor=south,align=center] at (-6,5.3) {Zoom A};
	
	\spy [orange, draw, height = 1cm, width = 2cm, magnification = 2] on (-5.2,3) in node at (-1.75,4.8);
	\node[anchor=south,align=center] at (-1.75,5.3) {Zoom B};
	
	\spy [orange, draw, height = 1cm, width = 2cm, magnification = 2] on (-.25,3.6) in node at (2.8,4.8);
	\node[anchor=south,align=center] at (2.8,5.3) {Zoom C};
	
	\spy [orange, draw, height = 1cm, width = 2cm, magnification = 2] on (5.5,2.75) in node at (6.9,4.8);
	\node[anchor=south,align=center] at (6.9,5.3) {Zoom D};
\end{tikzpicture}
\caption{One trajectory of a collateral agreement with rating triggers. The top picture shows the collateral account and portfolio over time, the middle one the rating evolution and the bottom one the corresponding rating thresholds.}%
\label{fig:collateralFOBBTwo}%
\end{figure}
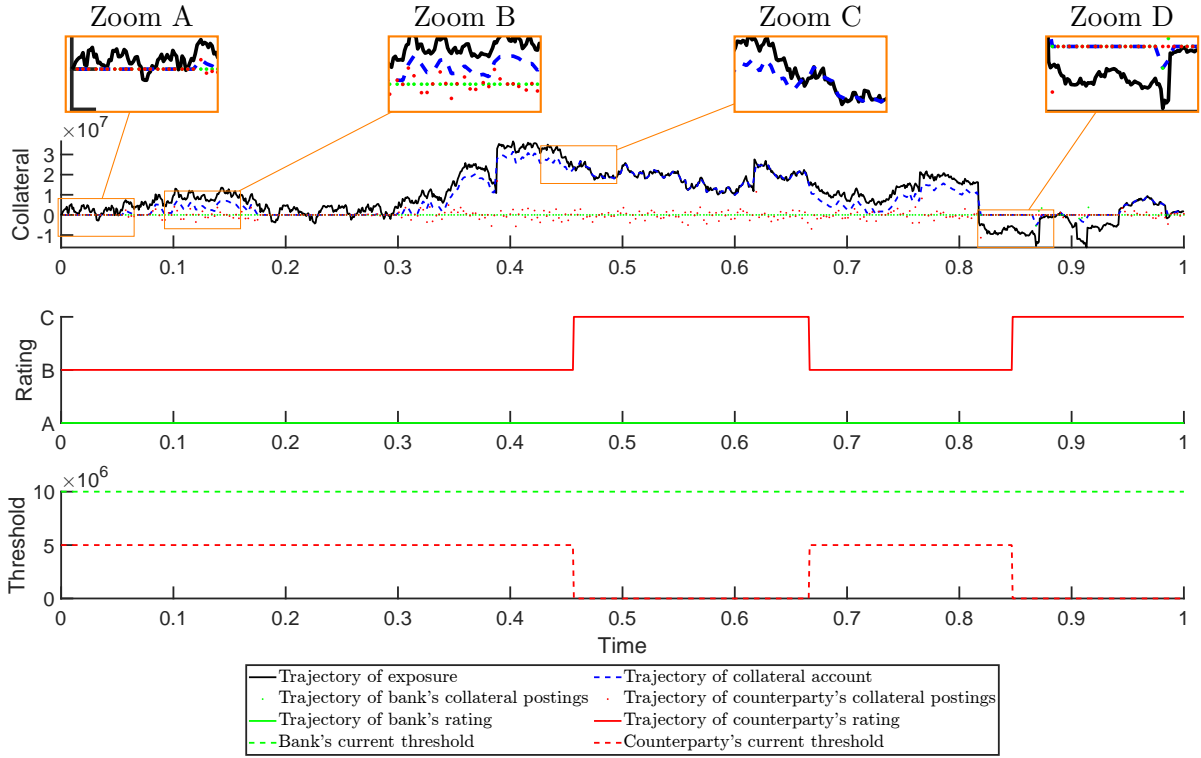

We will use 365 posting dates per year in all experiments.
In Figure \ref{fig:collateralFOBBTwo}, 
one can see one trajectory of the portfolio, collateral account and individual postings by both counterparties in the top picture. The picture in the middle indicates the ratings of both counterparties over time for this particular trajectory and the bottom picture shows the corresponding threshold for each point in time. The orange boxes are magnifications of the indicated sections in the figure.

One can see that for this choice of trajectory, the bank has no rating transition and the counterparty has many, ranging through all thresholds, which we set to $10$ million Euros for rating \rFormat{A}, 5 million Euros for \rFormat{B}, and zero for \rFormat{C} to force a perfect collateralization in this rating.

At the section ``Zoom A'', one can see that neither the bank nor the counterparty has to post collateral. In the region, where the portfolio is negative, the bank does not have to post collateral, since the allowed threshold is not exceeded. Same for the counterparty in the region, where the portfolio is positive. In section ``Zoom B'', the portfolio gets too positive and the threshold for the counterparty is exceeded, such that the counterparty has to post collateral (red dots). After that the portfolio gets negative again and oscillates around zero, such that no one has to post collateral again.
At ``Zoom C'', the rating of the counterparty drops to \rFormat{C}, forcing the perfectly collateralized scenario, since the threshold is zero for this rating. Therefore, the dashed blue line follows the black bold line perfectly.
Around $t=0.65$ the rating of the counterparty improves to \rFormat{B} permitting again a certain amount of unsecured money.
At ``Zoom D'', the ratings drops back to \rFormat{C}, which has no impact at this moment, since the exposure is negative. At the very end, the collateral follows the exposure again, whenever its above zero and would mitigate a large portion of potential losses due to a default event. 


In Table \ref{tab:CXVAFOBB},
are the values of \textCDVA \eqref{eq:CDVA}, \textCCVA \eqref{eq:CCVA} and \textCBVA \eqref{eq:CBVA} using
the Loss-Given-Default
$\lgd_B=0.6$, $\lgd_C=0.6$ and $M=10000$ simulations for 
the three collateral agreements: no collateralization, perfect collateralization and collateralization with rating triggers.

One can see that the collateralization with rating triggers lies in between the values of the uncollateralized case and the perfectly collateralized case, which is the expected behaviour, because as illustrated in Figure \ref{fig:collateralFOBBTwo} one has a possible transition from unsecured money to the perfectly collateralized scenario, where rating thresholds are zero. The difference to the perfectly collateralized case is that there can be transitions from high ratings to default in one instant, which will be subject of the next subsection.

\begin{table}%
\centering
\caption{\textCXVA with the different collateral agreements (no, perfectly and rating triggers) using case 2 in \Cref{tab:defaultProbabilities} as market default probabilities and $\lgd_B=0.6$, as well as $\lgd_C=0.6$ 
 with  $M=10000$ simulations.}
\begin{tabular}{|l|*{3}{c}|}
	\hline
	\textCXVA		& Uncollateralized 	 & Rating Triggers 		& Perfectly collateralized\\
	\hline
	\textCDVA 		& $246444$ & $131789$ & $83355$\\
	\textCCVA		& $195728$ & $87322$ & $78635$\\
	\textCBVA 		& $50715$ & $44466$ & $4719$\\
	\hline
\end{tabular}
\label{tab:CXVAFOBB}
\end{table}

\begin{remark}\label{rem:twoSectors}%
	In this framework of rating transition modelling, it is straightforward to include the possibility of counterparties in two different sectors, e.g., financial and corporate.
	
	Suppose we are in the setting of \Cref{sec:LieCalibrationQ} and define two independent processes in the Lie algebra, whose components are 
	\begin{align*}
		d\generator_t^{1,i} &= \abs{\generator_t^{1,i}}^{a_i^1} dt, \quad
		dY_t^{1,i}= b_i^1 dt + \sigma_i^1 dW_t^{1,i},\\
		d\generator_t^{2,i} &= \abs{\generator_t^{2,i}}^{a_i^2} dt, \quad
		dY_t^{2,i}= b_i^2 dt + \sigma_i^2 dW_t^{2,i}.
	\end{align*}
	The Brownian motions $W_t^{j,i}$, $i=1,\dots,(K-1)^2$, $j=1,2$, are assumed to be mutually independent.
	
	Now, instead of defining two individual changes of measures for both processes, we will define $\kappa \coloneqq \left(\kappa^1,\kappa^2\right) \in \R^{2(K-1)^2}$ and $W_t \coloneqq \left(W_t^{1,1},\dots,W_t^{1,(K-1)^2},W_t^{2,1},\dots,W_t^{2,(K-1)^2}\right)^\top$. The Girsanov transform now takes again the form
	\begin{align*}
		L_t\coloneqq \exp\left(\int_{0}^{t}{\kappa_s \cdot dW_s}-\frac{1}{2}\int_{0}^{t}{\abs{\kappa_s}^2 ds}\right)
	\end{align*}
	and the corresponding measure is given by 
	\begin{align*}
		\left.
			\frac{d\Q^\kappa}{d\P}
		\right|_{\mathcal{F}_t}\coloneqq L_t,
	\end{align*}
	where $\mathcal{F}_t \coloneqq \sigma\left(W_t\right)$ with $\Q^\kappa$ Brownian motion 
	$W_t^\kappa \coloneqq W_t-\int_{0}^{t}{\kappa_s ds}$.
	
	The dynamics of $\generator_t^1$ and $\generator_t^2$ under this new measure are given by
	\begin{align*}
		d\generator_t^{\kappa,1,i} &= \abs{\generator_t^{\kappa,1,i}}^{a_i^1} dt, \quad
		dY_t^{\kappa,1,i}= \left(b_i^1+\sigma_i^1\kappa_s^{1,i}\right) dt + \sigma_i^1 dW_t^{\kappa,1,i},\\
		d\generator_t^{\kappa,2,i} &= \abs{\generator_t^{\kappa,2,i}}^{a_i^2} dt, \quad
		dY_t^{\kappa,2,i}= \left(b_i^2+\sigma_i^2\kappa_s^{2,i}\right) dt + \sigma_i^2 dW_t^{\kappa,2,i}.
	\end{align*}
	This means that we can repeat the calibration procedure described in \Cref{sec:LieCalibrationP} and \Cref{sec:LieCalibrationQ} simultaneously for both processes using data from two different sectors, while ensuring that they have dynamics under the same risk-neutral measure $\Q^\kappa$.
	
\end{remark}

\subsection{Pre-Default Rating Distribution}\label{sec:preDefault}
\begin{figure}%
\centering
\includegraphics[width=.49\columnwidth]{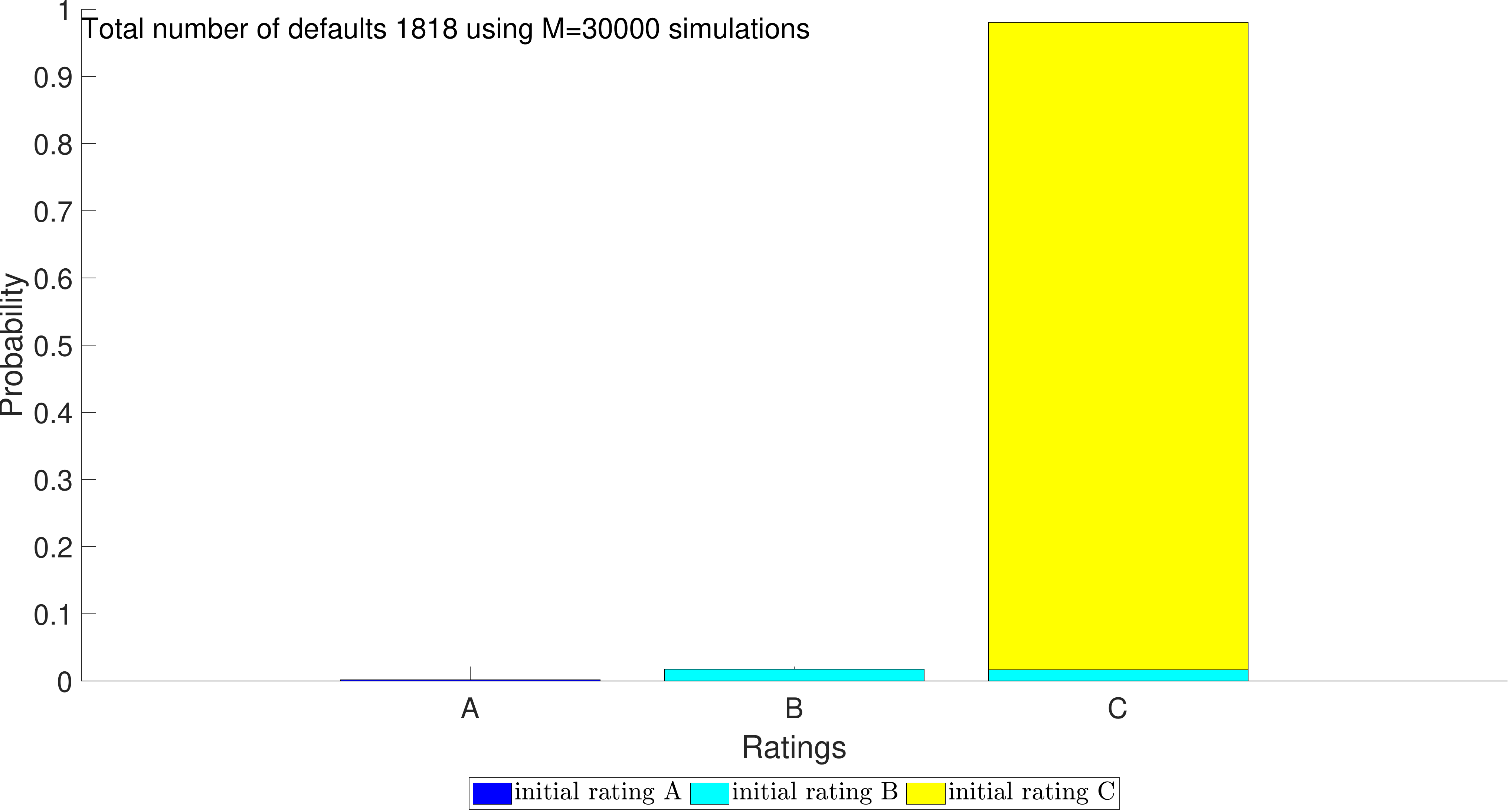}
\includegraphics[width=.49\columnwidth]{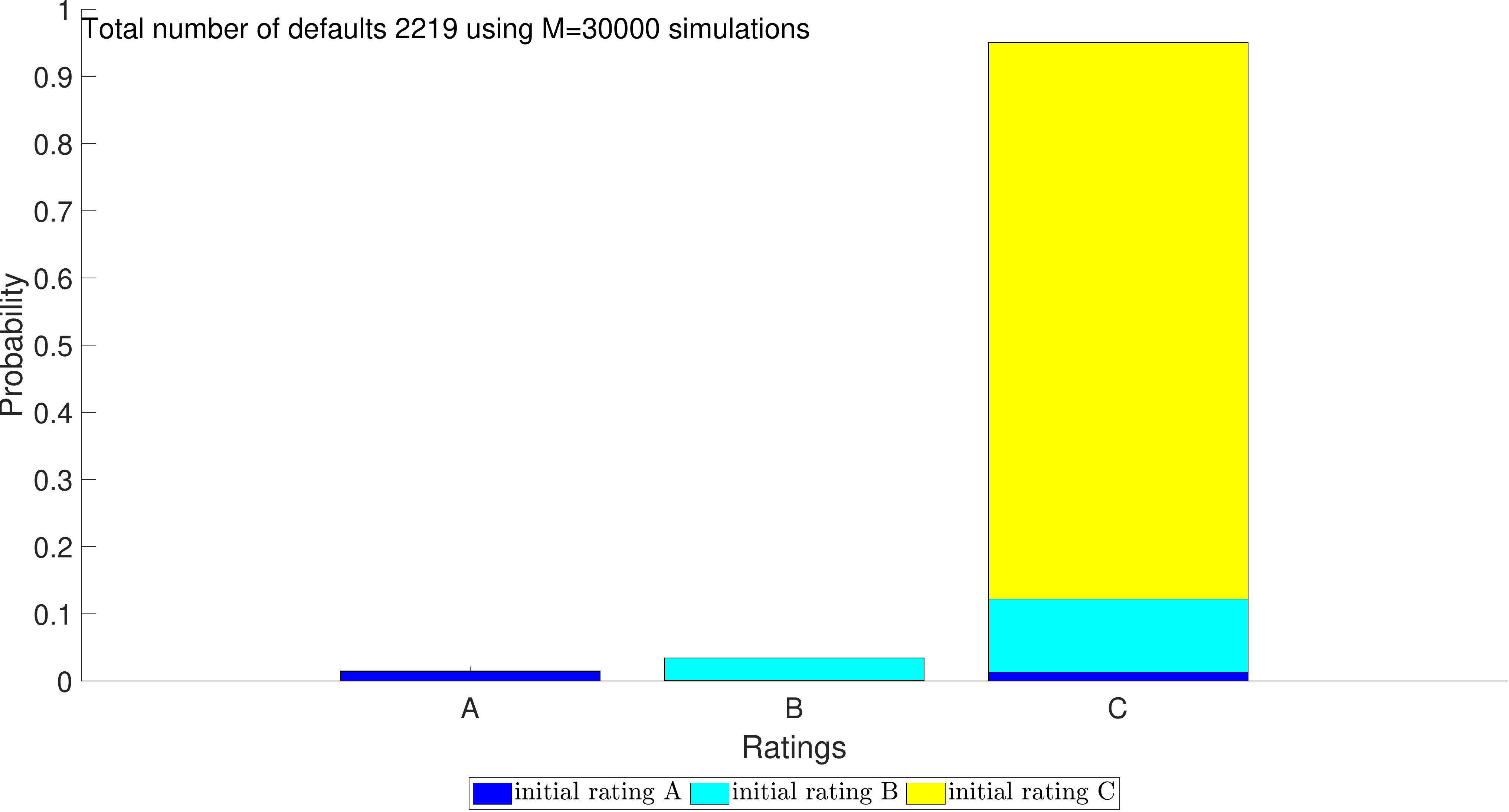}
\caption{Pre-default distribution using $\rGEM_t$. The left picture is under the historical measure and the right plot under the risk-neutral measure using the exponential change of measure with case 2 in \Cref{tab:defaultProbabilities} as market default probabilities.}%
\label{fig:PrePDFOBBSeven}%
\end{figure}

The definition of the rating thresholds motivates the necessity of studying the distribution of the rating process one time instant prior to default since it will determine the unsecured amount of money at the default event.
We will call this henceforth pre-default distribution and will also compare the distribution under $\P$ to the one under $\Q$ with the help of Figure \ref{fig:PrePDFOBBSeven}, which were obtained by Monte-Carlo simulation.

Now, let us have a closer look at Figure \ref{fig:PrePDFOBBSeven}. 
First of all, one can see the pre-default distribution
under the measure $\P$ in the top picture and under the measure $\Q$ in the bottom picture.
Disregarding the individual colors, the probability of being in a certain rating prior to default is given by the total height of the column. 
The composition of the individual colors of each column indicates the contribution of each starting rating, e.g., in the third column we can see that the most prominent contributions are resulting from the initial rating \rFormat{C}, but there are contributions of the other ratings as well.

In the market, it can be observed that the default probabilities in the risk-neutral world are usually higher than the default probabilities quoted under the historical measure in the rating matrices. This phenomenon has an impact in our model on all other ratings as well, which can already be seen in \Cref{fig:GEMsimulation1} 
by the spread of the grey lines indicating all simulated trajectories. In the risk-neutral world, there seem to be more transitions than in the historical world causing this spread of grey lines. 
The reason for this is that the calibration of this model has essentially one parameter for each rating because $h_i\in \R^K$. Therefore, the higher probability of default under the measure $\Q$ compared to the one under $\P$ has a significant impact on the other ratings as well.

We can see that under the measure $\P$, the top picture, almost all the defaults had a prior rating of \rFormat{C}, while under the measure $\Q$, this is still the most prominent case but significantly smaller. It is more likely under the measure $\Q$ that a company starting with a high rating defaults and this without transitioning to the rating prior to default first, which is indicated by the different heights of the each individual color for each rating.

It is yet an open question and needs thorough economical investigation whether this behaviour makes sense or not, because it has a significant impact on the performance of collateralization with rating triggers. To be more precise, the more likely it is that a company starting in a good rating defaults without first transitioning to a rating, where a low threshold is defined, the more unsecured money we have at a default event.

\section{Conclusion and Future Research}\label{sec:Conclusion}
In this paper, we proposed a novel rating transition model using Lie group techniques, which is fully inhomogeneous and stochastic. We demonstrated how to solve this SDE numerically by using the geometric Euler-Maruyama scheme and how to simulate rating processes by using a nested stochastic simulation algorithm. We applied these rating processes to price \textCXVA with rating triggers. Thus, it was crucial to apply Girsanov's theorem to account for both historical and risk-neutral data. We have compared two popular change of measure techniques and found that the exponential change of measure outperformed the JLT change of measure in our test cases. For the historical data, we proposed a novel reconstruction technique for cohort rating matrices. We used the trained Autoencoder of a TimeGAN to remove effects due to withdrawals. Moreover, we accounted for uncertainty using this approach during the calibration of our stochastic rating transition model.

As mentioned in \Cref{sec:ratingCohort} and \Cref{sec:LieCalibrationP}, we would like to change the procedure of estimating the reconstruction uncertainty in the future. In this respect, we would like to change the DNN in such a way, that it also estimates its own uncertainty in the reconstruction.

Another line of research concerns the volatility parameter as well. We would like to use our rating transition model as a doubly stochastic default model and calibrate it to CDS quotes. This would allow us to extract information about the market volatilities similar to a calibration to zero-coupon prices in an interest rate framework (cf. \cite{DFK2021} and \cite{DFK2022}).

\appendix

\section*{Declarations}
\subsection*{Funding}
This project has received funding from the European Union’s Horizon 2020 research and innovation
programme under the Marie Sklodowska-Curie grant agreement No 813261 and is part of the ABC-EU-XVA project.
\subsection*{Conflicts of interests}

The authors have no relevant financial or non-financial interests to disclose.

\subsection*{Data availability}
All data generated or analysed during this study are included in this published article except 
for the historical rating transition data, which has to be downloaded from the respective websites
of the rating agencies while agreeing to their terms of usage.
The code and data sets to produce the numerical experiments are available at
\url{https://github.com/kevinkamm/StochasticCohort}.
{\thispagestyle{scrheadings}
\thispagestyle{scrheadings}\ihead{}
\singlespacing
\begin{footnotesize}
\bibliographystyle{acm}
\bibliography{literature.bib}

\begin{thebibliography}{10}

\bibitem{Arns2010}
{\sc Arns, M., Buchholz, P., and Panchenko, A.}
\newblock On the numerical analysis of inhomogeneous continuous-time markov
  chains.
\newblock {\em INFORMS Journal on Computing 22}, 3 (2010), 416--432.

\bibitem{Bielecki2012}
{\sc Bielecki, T., Cialenco, I., and Iyigunler, I.}
\newblock Collateralized cva valuation with rating triggers and credit
  migrations.
\newblock {\em International Journal of Theoretical and Applied Finance 16\/}
  (05 2012).

\bibitem{Brigo2013}
{\sc Brigo, D., Morini, M., and Pallavicini, A.}
\newblock {\em Counterparty Credit Risk, Collateral and Funding: With Pricing
  Cases For All Asset Classes}.
\newblock The Wiley Finance Series. Wiley, 2013.

\bibitem{Coletti2020}
{\sc Coletti, C., Carneiro, R., and Yepes, S.}
\newblock Some geometric properties of stochastic matrices.
\newblock {\em Proceeding Series of the Brazilian Society of Computational and
  Applied Mathematics 7}, 1 (02 2020).

\bibitem{DFK2021}
{\sc Di~Francesco, M., and Kamm, K.}
\newblock How to handle negative interest rates in a cir framework.
\newblock {\em SeMA Journal\/} (Oct 2021).

\bibitem{DFK2022}
{\sc Di~Francesco, M., and Kamm, K.}
\newblock On the deterministic-shift extended cir model in a negative interest
  rate framework.
\newblock {\em International Journal of Financial Studies 10}, 2 (2022).

\bibitem{Gillespie2007}
{\sc Gillespie, D.~T.}
\newblock Stochastic simulation of chemical kinetics.
\newblock {\em Annual Review of Physical Chemistry 58}, 1 (2007), 35--55.
\newblock PMID: 17037977.

\bibitem{HairerLubichWanner}
{\sc Hairer, E., Lubich, C., and Wanner, G.}
\newblock {\em Geometric numerical integration}, second~ed., vol.~31 of {\em
  Springer Series in Computational Mathematics}.
\newblock Springer-Verlag, Berlin, 2006.
\newblock Structure-preserving algorithms for ordinary differential equations.

\bibitem{Hall03}
{\sc Hall, B.}
\newblock {\em Lie Groups, Lie Algebras, and Representations: An Elementary
  Introduction}.
\newblock Graduate Texts in Mathematics. Springer, 2003.

\bibitem{Israel2001}
{\sc Israel, R.~B., Rosenthal, J.~S., and Wei, J.~Z.}
\newblock Finding generators for markov chains via empirical transition
  matrices, with applications to credit ratings.
\newblock {\em Mathematical Finance 11}, 2 (2001), 245--265.

\bibitem{Jarrow1997}
{\sc Jarrow, R., Lando, D., and Turnbull, S.~M.}
\newblock A markov model for the term structure of credit risk spreads.
\newblock {\em Review of Financial Studies 10}, 2 (1997), 481--523.

\bibitem{K2022}
{\sc Kamm, K.}
\newblock An introduction to rating triggers for collateral-inclusive xva in an
  ictmc framework.

\bibitem{KM2022}
{\sc Kamm, K., and Muniz, M.}
\newblock A novel approach to rating transition modelling via machine learning
  and sdes on lie groups.

\bibitem{KPP2021}
{\sc Kamm, K., Pagliarani, S., and Pascucci, A.}
\newblock On the stochastic magnus expansion and its application to spdes.
\newblock {\em Journal of Scientific Computing 89}, 3 (Oct 2021), 56.

\bibitem{Kingma2014}
{\sc Kingma, D.~P., and Ba, J.}
\newblock Adam: A method for stochastic optimization, 2014.

\bibitem{Lando2002}
{\sc Lando, D., and Sk{\o}deberg, T.~M.}
\newblock Analyzing rating transitions and rating drift with continuous
  observations.
\newblock {\em Journal of Banking {\&} Finance 26}, 2 (2002), 423--444.

\bibitem{Lencastre2014}
{\sc Lencastre, P., Raischel, F., Lind, P.~G., and Rogers, T.}
\newblock Are credit ratings time-homogeneous and markov?, 2014.

\bibitem{Li2012}
{\sc Li, Y.-F., Lin, Y.-H., and Zio, E.}
\newblock {Stochastic Modeling by Inhomogeneous Continuous Time Markov Chains}.

\bibitem{LordMalhamSimonWiese}
{\sc Lord, G., Malham, S. J.~A., and Wiese, A.}
\newblock Efficient strong integrators for linear stochastic systems.
\newblock {\em SIAM J. Numer. Anal. 46}, 6 (2008), 2892--2919.

\bibitem{Lu2019}
{\sc Lu, L., Jin, P., and Karniadakis, G.~E.}
\newblock Deeponet: Learning nonlinear operators for identifying differential
  equations based on the universal approximation theorem of operators.
\newblock {\em CoRR abs/1910.03193\/} (2019).

\bibitem{GoranSolo}
{\sc Marjanovic, G., and Solo, V.}
\newblock Numerical methods for stochastic differential equations in matrix
  {L}ie groups made simple.
\newblock {\em IEEE Trans. Automat. Control 63}, 12 (2018), 4035--4050.

\bibitem{Muniz2021}
{\sc Muniz, M., Ehrhardt, M., G\"{u}nther, M., and Winkler, R.}
\newblock Higher strong order methods for linear {I}t\^o {SDE}s on matrix {L}ie
  groups.
\newblock {\em To appear in BIT Numer Math\/} (2022),
  https://doi.org/10.1007/s10543--022--00911--5.

\bibitem{Muniz2022}
{\sc Muniz, M., Ehrhardt, M., Günther, M., and Winkler, R.}
\newblock Strong stochastic runge-kutta-munthe-kaas methods for nonlinear it\^o
  sdes on manifolds, 06 2022.

\bibitem{MuntheKaas1999}
{\sc Munthe-Kaas, H.}
\newblock High order runge-kutta methods on manifolds.
\newblock {\em Applied Numerical Mathematics 29}, 1 (1999), 115--127.
\newblock Proceedings of the NSF/CBMS Regional Conference on Numerical Analysis
  of Hamiltonian Differential Equations.

\bibitem{Oosterlee2019}
{\sc Oosterlee, C., and Grzelak, L.}
\newblock {\em Mathematical Modeling And Computation In Finance: With Exercises
  And Python And Matlab Computer Codes}.
\newblock World Scientific Publishing Company, 2019.

\bibitem{Palmowski2002}
{\sc Palmowski, Z., and Rolski, T.}
\newblock {A technique for exponential change of measure for Markov processes}.
\newblock {\em Bernoulli 8}, 6 (2002), 767 -- 785.

\bibitem{Stroock2005}
{\sc Stroock, D.~W.}
\newblock {\em An Introduction to Markov Processes}, 1~ed.
\newblock Graduate Texts in Mathematics. Springer, Heidelberg, 2005.

\bibitem{Yoon2019}
{\sc Yoon, J., Jarrett, D., and van~der Schaar, M.}
\newblock Time-series generative adversarial networks.
\newblock In {\em Advances in Neural Information Processing Systems\/} (2019),
  H.~Wallach, H.~Larochelle, A.~Beygelzimer, F.~d\textquotesingle
  Alch\'{e}-Buc, E.~Fox, and R.~Garnett, Eds., vol.~32, Curran Associates, Inc.

\end{thebibliography}
\end{footnotesize}
}
\end{document}